\documentclass[a4paper,fleqn,usenatbib]{mnras}

\usepackage[T1]{fontenc}
\usepackage{ae,aecompl}

\usepackage{graphicx}	
\usepackage{amsmath}	
\usepackage{amssymb}	

\title{High-redshift major mergers weakly enhance star formation}
\author[J. Fensch et al.]{J. Fensch$^{1}$\thanks{E-mail: jeremy.fensch@cea.fr}, 
F. Renaud$^{2}$, F. Bournaud$^{1}$, P.-A. Duc$^{1}$, O. Agertz$^{2}$, P. Amram$^{3}$, \newauthor F. Combes$^{4}$, P. Di Matteo$^{5}$, B. Elmegreen$^{6}$, E. Emsellem$^{7,8}$, C. J. Jog$^{9}$, V. Perret$^{10}$, \newauthor C. Struck$^{11}$ \& R. Teyssier$^{10}$ \\
\\
$^{1}$ Laboratoire AIM Paris-Saclay, CEA/IRFU/SAp, CNRS, Universite Paris Diderot, F-91191 Gif-sur-Yvette Cedex, France\\
$^{2}$ Department of Physics, University of Surrey, Guildford, GU2 7XH, UK\\
$^{3}$Universite Aix Marseille , CNRS, LAM (Laboratoire d'Astrophysique de Marseille), 13388, Marseille, France\\
$^{4}$ Observatoire de Paris, LERMA (CNRS: UMR 8112), 61 Av. de l'Observatoire, 75014 Paris, France\\
$^{5}$ GEPI, Observatoire de Paris, PSL Research University, CNRS, Univ Paris Diderot, Sorbonne Paris Cité,\\ Place Jules Janssen, 92195 Meudon, France\\
$^{6}$ IBM Research Division, T.J. Watson Research Center, P.O. Box 218, Yorktown Heights, NY 10598, USA\\
$^{7}$ European Southern Observatory, D-85748 Garching bei Munchen, Germany\\
$^{8}$ Universite Lyon 1, Observatoire de Lyon, CRAL et ENS, 9 Av Charles Andre, F-69230 Saint-Genis Laval, France\\
$^{9}$ Department of Physics, Indian Institute of Science, Bangalore 560012, India\\
$^{10}$ Institute for Theoretical Physics, University of Zurich, CH-8057 Zurich, Switzerland\\
$^{11}$ Department of Physics and Astronomy, Iowa State University, Ames, IA 50014 USA\\
}

\date{Accepted XXX. Received YYY; in original form ZZZ}

\pubyear{2016}

\begin{document}
\label{firstpage}
\pagerange{\pageref{firstpage}--\pageref{lastpage}}
\maketitle

\begin{abstract}

Galaxy mergers are believed to trigger strong starbursts. This is well assessed by observations in the local Universe. However the efficiency of this mechanism has poorly been tested so far for high redshift, actively star forming, galaxies. We present a suite of pc-resolution hydrodynamical numerical simulations to compare the star formation process along a merging sequence of high and low $z$ galaxies, by varying the gas mass fraction between the two models. We show that, for the same orbit, high-redshift gas-rich mergers are less efficient than low-redshift ones at producing starbursts: the star formation rate excess induced by the merger and its duration are both around 10 times lower than in the low gas fraction case.
The mechanisms that account for the star formation triggering at low redshift - the increased compressive turbulence, gas fragmentation, and central gas inflows - are only mildly, if not at all, enhanced for high gas fraction galaxy encounters. Furthermore, we show that the strong stellar feedback from the initially high star formation rate in high redshift galaxies does not prevent an increase of the star formation during the merger. Our results are consistent with the observed increase of the number of major mergers with increasing redshift being faster than the respective increase in the number of starburst galaxies.

\end{abstract}




\section{Introduction}

Observations of star forming galaxies show that they follow a tight correlation between their stellar mass ($M_{\star}$) and their star formation rate (SFR). This relation defines a main sequence (MS) that is observed over a wide redshift range ($z=4$ to 0) \citep{Elbaz07, Noeske07, Peng10, Rodighiero11, Schreiber15}.

Outlier galaxies, which display higher specific star formation rate (sSFR) than members of the MS, correspond to the starbursting galaxies. Detailed observations of these galaxies in the local Universe show that, above a certain luminosity threshold ($L_{\mathrm{IR}} > 10^{12}$ L$_{\odot}$, which defines the ultra luminous infrared galaxies, ULIRGs, see e.g. \citealt{Houck85}), all galaxies are undergoing a major merger event \citep{Armus87, Sanders96, Ellison13}. It should also be noted that if all local starburst galaxies are undergoing an interaction it is not reciprocal, as all observed major mergers do not show a massive enhancement of star formation (at low redshift, see \citealt{Bergvall03}, at high redshift, see \citealt{Jogee09}). This might however be due to a short extent of the starburst phase relative to the interaction timescale.

Observations of star-forming galaxies have shown that the number of star forming galaxies in the starbursting regime above the MS remains fairly constant (between 2\% and 4 \%) in the redshift range $z=4$ to $z=0$, with no or only weak variation with redshift: about twice at most between $z=0$ and $z=2$ \citep{Rodighiero11, Schreiber15}. Given that the major merger rate is thought to be an increasing function of redshift \citep[][and references therein]{LeFevre00, Kampczyk07, Kartaltepe07, Lotz11}, this implies that the efficiency of major mergers in triggering starbursts must decrease with increasing redshift. This hypothesis is backed by observations showing that major mergers are inefficient at driving star formation at $z \simeq 2$ \citep{Kaviraj13, Lofthouse16} and calls for a detailed study of physical processes at play in galaxy major mergers, and their dependance on redshift.

Observations of local advanced mergers show that most of the star formation is concentrated in the inner regions \citep{Sanders96, Duc97}. These nuclear starbursts are explained by gravitational torques induced by the close interaction between the galaxies \citep{Barnes91}. However, gravitationally driven gas inflows alone cannot explain observations of extended, off-nuclear increase of interaction-induced star formation \citep{Barnes04, Chien10, Smith14}. One the most striking case if the well-studied Antennae system, which is observed during the second pericentre passage \citep{Renaud08} and where the majority of star formation happens in dense clumps outside of the nuclei and in the overlap region \citep[see e.g.][]{Whitmore95}. Atomic and molecular gas observations \citep[such as][for the NGC~2207 -- IC~2163 interacting pair]{Elmegreen16}, show that their star forming regions have very dense molecular phases, which hints for strong gas compression.

Massive high-redshift ($z>2$) galaxy disks are characterized by a high gas fraction: $f_{\mathrm{gas}} \simeq 50\%$ at $z=2$ to be compared with $f_{\mathrm{gas}} \simeq 10\%$ at $z=0$ \citep{Daddi10a, Tacconi10}. This higher gas fraction makes the disk more prone to gravitational instabilities: the velocity dispersion is higher ($\sigma \simeq$ 40 km/s at $z=2$, \citealt{Forster09, Stott16, Price16}, to be compared with $\sigma \simeq$ 10 km/s at $z=0$) and the violent disk instability (VDI) induces strong nuclear inflows \citep{Bournaud12} and the formation of giant star forming clumps of $10^{ 7}$ to $10^{9}$ M$_{\odot}$\citep{Elmegreen07, Elmegreen09b, Genzel08, Magnelli12, Guo15, Zanella15}. At this redshift, it is hard to study the evolution of the galaxy morphology during encounters \citep[see e.g.][]{Cibinel15}, and especially the efficiency of gas fragmentation and nuclear inflows which trigger the starburst at low redshift.

Numerical simulations are particularly well-suited to capture the complex and rapidly evolving dynamics of mergers. It has been shown that the spatial resolution plays an important role in the study of extended star formation \citep{DiMatteo08, Teyssier10, Bournaud11}: one should be able to resolve the local Jeans length with several resolution elements  to be able to capture the off-nuclei gas condensations that might also form stars. Merger simulations with resolution of a few 10~pc to pc-scale \citep [ R14 in the following]{Teyssier10, Powell13, Renaud14} yield an extended star-formation, which is explained by an increase of the cold gas turbulence, and especially of its compressive mode, driven by interaction-induced fully compressive tides \citep{Renaud08, Renaud09}. This process adds to the torque-driven gas inflows towards the nuclei in the enhancement of the star formation rate (SFR). Fragmentation of the gas is relatively more important than nuclear inflows at the first pericentre passages, when the impact parameter and the relative velocity are large, resulting in weak and short-lived gravitational torques. 

Previous simulations of galaxy major mergers with a high gas fraction hint for a weak enhancement of star formation, both in burst amplitude and duration \citep[see e.g.][]{Bournaud11, Hopkins13, Scudder15}, or even not at all \citep{Perret14}. However, they did not compare their sample to low gas fraction major mergers or investigate the physical cause of this mild burst.

In this paper, we present a comparison between low and high redshift major mergers using hydrodynamical simulations. We present two sets of simulations: one with a high gas fraction, and one with a low gas fraction. We study the evolution of the properties of their star formation activity during the interactions. Section 2 presents the numerical simulation parameters. Section 3 describes the resulting evolution of the star formation activity during the interactions, and compare the high and low gas fraction cases. The analysis of the physical processes having a role in the star formation is presented in Section 4. Discussion and conclusions are drawn in Sections 5 and 6.


\section{Simulations}

 \subsection{Numerical technique}
 \label{code}

We perform the simulations using the adaptative mesh refinement (AMR) code \textsc{ramses} \citep{Teyssier02}. The overall method is analogous of that in \citet{Renaud15}.   

The coarsest grid is composed of $64^{3}$ cells in a cubic box of 200~kpc, and we allow up to 10 further levels of refinement. This results in a maximum spatial resolution of 6~pc. A resolution study to discuss this choice is presented in Appendix~\ref{resolution}. Star and dark matter (DM) particles are also implemented in the initial conditions (see Section~\ref{model}). We call the former old stars, to differentiate them from the new stars, that are formed during the simulation. A grid cell is refined if when there are more than 40 initial condition particles, or if the baryonic mass, including gas, old and new stars, exceeds $8\times10^{5}M_{\odot}$. Furthermore, we ensure that the Jeans length is always resolved by at least four cells, by introducing a numerical pressure through a temperature floor set by a polytrope equation of state at high density ($T \propto \rho^{2}$), which is called Jeans polytrope thereafter, which prevents numerical fragmentation \citep{Truelove97}. Such a polytrope was also added in numerous previous numerical simulations \citep{Dubois08, Renaud13, Bournaud14, Kraljic14, Perret14}. 

The thermodynamical model, including heating and atomic cooling at solar metallicity\footnote{We use the same metallicity for both high and low redshift galaxy models as an approximation to investigate the effects imputable to the variation of gas fraction only. Massive high redshift galaxies can have a metallicity up to half-solar \citep[see e.g.][]{Erb06}.}, is the same as described in \citet{Renaud15}. 

If the gas reaches a certain density threshold, $\rho_{0}$, and if its temperature is no more than $2\times10^{4}$~K above the Jeans polytrope temperature at the corresponding density, it is converted into stellar particles following a Schmidt law, $\dot{\rho}_{\star} = \epsilon (\rho_{\mathrm{gas}} / t_{\mathrm{ff}})$ \citep{Schmidt59, Kennicutt98}, with $\epsilon$ the efficiency per free fall time set to 1\%, and $ t_{\mathrm{ff}} = \sqrt{3\pi / (32G\rho)}$ the free-fall time. The density thresholds ($\rho_{0}$ = 30 cm$^{-3}$ and $\rho_{0}$ = 1e5 cm$^{-3}$ for low and high gas fraction cases) are chosen to correspond to the isolated  disks being on the disk sequence of the Schmidt-Kennicutt diagram \citep{Daddi10a, Genzel10}, that is a star formation rate (SFR) of $ \simeq 1~$M$_{\odot}$/yr (resp. $ \simeq 60~$M$_{\odot}$/yr) in the low gas fraction case (resp. high gas fraction case) for each galaxy before the interaction. The difference in the normalization of $\rho_{0}$ originates from the different gas mean density between the two cases\footnote{A density threshold closer to resolved highest density will make the SFR more sensitive to small variations in the shape the density distribution. This is the case for our gas-rich case (see Section~\ref{sectionPDF}), and it strengthens our results (see Section~\ref{sb}).}.\\

\begin{table} 
\centering
\caption{Characteristics of the galaxies used in the simulations.\label{galaxy}}
\begin{tabular}{|p{5.25cm}||p{1.05cm}||p{1.05cm}|}
\textbf{Galaxy}  & low gas fraction & high gas fraction \\ \hline \hline
 
Total baryonic mass {[} $\times 10^{9} $M$_{\odot} ${]} & \multicolumn{2}{c}{57.2}                  \\ \hline

\textbf{Gas Disc (exponential profile)}           &               & \multicolumn{1}{l}{}     \\
mass  {[} $\times 10^{9} $M$_{\odot} ${]}   &    \multicolumn{1}{c}{5.0} &    \multicolumn{1}{c}{34.3}\\
characteristic radius {[}kpc{]}           & \multicolumn{2}{c}{8.0}                   \\
truncation radius {[}kpc{]}             & \multicolumn{2}{c}{14.0}                  \\
characteristic height {[}kpc{]}           & \multicolumn{2}{c}{0.3}                   \\
truncation height {[}kpc{]}             & \multicolumn{2}{c}{0.8}                   \\ \hline

\textbf{Stellar Disk (exponential profile)}         & \multicolumn{1}{l}{}     & \multicolumn{1}{l}{}     \\
number of particles             &    \multicolumn{1}{c}{500~000} &  \multicolumn{1}{c}{173~900}                    \\
mass  {[} $\times 10^{9} $M$_{\odot} ${]}   & \multicolumn{1}{c}{45.0}  & \multicolumn{1}{c}{15.7}                    \\
characteristic radius {[}kpc{]}           & \multicolumn{2}{c}{5.0}                    \\
truncation radius {[}kpc{]}             & \multicolumn{2}{c}{12.0}                    \\
characteristic height {[}kpc{]}           & \multicolumn{2}{c}{0.34}                    \\
truncation height {[}kpc{]}             & \multicolumn{2}{c}{1.02}                    \\ 
\hline

\textbf{Bulge (Hernquist profile)}              & \multicolumn{1}{l}{}     & \multicolumn{1}{l}{}     \\
number of particles                 & \multicolumn{2}{c}{80~000 }                    \\
mass   {[} $\times 10^{9} $M$_{\odot} ${]}                  & \multicolumn{2}{c}{7.2 } \\
characteristic radius {[}kpc{]}           & \multicolumn{2}{c}{1.3}                    \\
truncation radius {[}kpc{]}             & \multicolumn{2}{c}{3.0}                    \\
 \hline
\textbf{Dark Matter Halo (Burkert profile)}              & \multicolumn{1}{l}{}     & \multicolumn{1}{l}{}     \\
number of particles                 & \multicolumn{2}{c}{ 500~000}                    \\
mass   {[} $\times 10^{9} $M$_{\odot} ${]}                   & \multicolumn{2}{c}{ 131.0 } \\   
characteristic radius {[}kpc{]}           & \multicolumn{2}{c}{25.0}                    \\
truncation radius {[}kpc{]}             & \multicolumn{2}{c}{45.0}                    \\	
\end{tabular}
\end{table}

Because of the spatial resolution and the computational cost it would imply, we do not resolve individual star formation. We chose a high enough sampling mass for the new stars, i.e higher than 1000~M$_{\odot}$, so that we do not need to resolve the initial mass function.

In these simulations we model three types of stellar feedback :
\begin{enumerate}
\item Photo-ionisation from H\textsc{ii} regions (as in \citealt{Renaud13}): UV photons from the OB-type stars ionize the surrounding gas. To model this process we evaluate the radius of the \citet{Stromgren39} sphere around the newly formed stellar particle and heat up the gas to $T_{\mathrm{ H\textsc{ii}}} = 2 \times 10^{4}$ K.

\item Radiation pressure (as in \citealt{Renaud13}): inside each H\textsc{ii} region the scattering of photons on the gas acts as a radiative pressure. To model this process we inject a radial velocity kick for each cell inside the Stromgren  sphere.
 
\item type-II supernova thermal blasts (as in \citealt{Dubois08} and \citealt{Teyssier13}): stars more massive than 4~M$_{\odot}$ eventually explode as type II supernovae \citep{Povich12}. We assume that 20\% of the initial mass of our stellar particle is in massive stars, and will be released 10~Myr after the formation of the stellar particle. On top of the mass loss, we inject thermally a certain amount of energy in the cell: $E_{\mathrm{SN}} = 10^{51}~$ergs$~/ ~10~$M$_{\odot}$.
\end{enumerate}
This implementation of stellar feedback is therefore physically motivated, although sub-grid. It has been used in numerical simulations of high gas fraction disk galaxies similar to ours, and was shown to have realistic effects on the galaxies, such as producing large scale outflows with a mass-loading factor close to unity (\citealt{Roos15}, and see Section~3.3 and Fig.~5 in \citealt{Bournaud16b}), which is consistent with observations \citep[see e.g.][]{Newman12}. 

\subsection{Modeling low and high-redshift galactic disks}
\label{model}

We run a suite of massive galaxy merger simulations for gas-rich ($f _{\mathrm{gas}}$ = $60\%$) and gas-poor disks ($f _{\mathrm{gas}}$ = 10$\%$). These gas fractions are typical of $z=2$ and $z=0$ disk galaxies. We chose to modify solely the gas fraction in order to isolate the effect of this parameter, and neglect the other differences between the low and high redshift, such as galaxy size, mass and interaction parameters. These parameters are discussed in Section~\ref{diff}. Furthermore, it must be noted that we do not change the stellar mass of the bulge, so that the velocity profile and the galactic shear stay the same, and so that the low and high gas fraction orbits are as similar as possible.

The DM halo for each galaxy is composed of $2.62\times10^{5}$~M$_{\odot}$ particles, following a \citet{Burkert95} profile. An old stellar component is added, made of $9 \times10^{4}$~M$_{\odot}$ particles modeling a stellar disk and a stellar bulge. The characteristics of the galaxies are summed up in Table~\ref{galaxy}. The DM, old stars and the stellar particles formed during the simulation are evolved through a particle-mesh solver, with a gravitational softening of 50~pc for the DM and old stars and at the resolution of the local resolution for the new stars. 

\subsection{Characteristics of the orbits}

\begin{table}
\centering

\caption{Initial conditions of the orbits \label{ic}}
\begin{tabular}{|p{2cm}||p{2.8cm}||p{2.8cm}|}
        & Galaxy 1        & Galaxy 2        \\ \hline 
        \hline
\textbf{Orbit \#1}     &             &              \\ 
Center {[}kpc{]}       & (10.55, -30.34, 46.68) & (-15.01, 30.44, -46.34)   \\
Velocity {[}km/s{]}     & (-26.95, 23.23, -71.76) & (26.02, -23.28, 71.35)  \\ \hline

\textbf{Orbit \#2 }          &             &              \\ 
Center {[}kpc{]}       & (8.64, -25.78, 39.71)  & (-13.10, 25.89, -39.36)   \\
Velocity {[}km/s{]}     & (-22.98, 19.74, -61.03) & (22.05, -19.80, 60.61)   \\ \hline

\textbf{Orbit \#3}          &             &             \\
Center {[}kpc{]}       & (6.21, -20.01, 30.87)  & (-10.67, 20.11, -30.52)  \\
Velocity {[}km/s{]}     & (-17.94, 15.32, -47.43) & (17.01, -15.37, 47.02)   \\ 

\end{tabular}\\
\end{table}

\begin{table} 
\centering
\caption{Orientations of the spin axis.\label{spins}}
\begin{tabular}{lcc}

\textbf{Spin} & Galaxy 1  & Galaxy 2 \\ \hline \hline

dd                    & (-0.67, -0.71, 0.20) &  (0.65, 0.65, -0.40)  \\ 
rr                     &  ( 0.67,  0.71, -0.20) &  (-0.65, -0.65, 0.40)  \\ \hline

spin1             &  ( -0.67,  0.71, -0.20) & (-0.65, 0.65, 0.40)    \\
spin2            &  ( 0.67, -0.71,  0.20) & (-0.65, -0.65, 0.40)   \\ 
spin3           &  ( 0.67, 0.71,  0.20) &( 0.65, -0.65, 0.40)  \\
spin4           &  (-0.67, 0.71,  -0.20) & (-0.65, -0.65, -0.40)   \\

\end{tabular}
\end{table}


\begin{table} 
\centering
\caption{Summary of the simulations. The prefix 'gp' stands for gas-poor. dd and rr stand for direct-direct and retrograde-retrograde, and are followed by the number of the corresponding orbit.\label{simulations}}
\begin{tabular}{|p{1cm}||p{1cm}||p{1cm}||p{1.75cm}||p{1.85cm}|}
\textbf{Name} & Orbit & Spins  & Gas fraction & Maximum resolution~[pc] \\ \hline \hline
iso & - & - & 60\% & 6 \\ 
gp-iso & - & - & 10 \% & 6 \\ \hline

dd1       &  1    & dd   & 60\%  & 6 \\
rr1       &  1    & rr     & 60\%  & 6 \\ 
gp-dd1       &  1    & dd  & 10\%  & 6 \\
gp-rr1       &  1    & rr    & 10\% & 6 \\ \hline

dd2       &  2    & dd     & 60\% & 12 \\
rr2       &  2    & rr    & 60\%  &12 \\ 
spin1      &  2    & spin1   & 60\%  &12 \\
spin2     &  2    & spin2  & 60\%  &12 \\
spin3      &  2    & spin3  & 60\%  &12 \\ 
spin4      &  2    & spin4 & 60\%  &12 \\ \hline

dd3       &  3    & dd     & 60\%  &12 \\
rr3       &  3    &  rr    & 60\%  &12 \\

\end{tabular}
\end{table}

Our simulation sample comprises three orbits.Their parameters are summarised in Tables~\ref{ic} and \ref{spins}. The full simulation suite is summarised in Table~\ref{simulations} . Low and high gas fraction mergers are run on the same reference orbit \#1 for comparison. This orbit is close to that of the Antennae system, which was shown by R14 to be favorable to a strong starburst for low gas fraction disks. Orbits \#2 and \#3 have lower orbital energy, obtained by reducing the impact parameter and relative velocity by 15\% and 30\%. This ensures that the results do not depend on an unforeseen particularity of the orbit. The initial orbital conditions are summarised in Table~\ref{ic}.

As the orientation of the galaxies plays a significant role in the processes at play \citep[see review by][]{Duc13}, we use a set of different spin vectors for each orbit (see Table~\ref{spins}). For each orbit we run one direct-direct, Antennae-like interaction and one retrograde-retrograde encounter. The latter is obtained by merely taking the opposite of the spin orientation of both galaxies. To ensure that the different spin-orbit couplings do not affect our conclusions, we also run 4 other simulations with a different spin orientation on orbit \#2.

We run orbit \#1 simulations at 6~pc resolution and the orbits \#2 and \#3 at 12~pc resolution, using the same refinement strategy as for the 6~pc case, stopped one level lower, and $\rho_{0} = $1e4 cm$^{-3}$ as star-formation threshold, which is lower than for the 6~pc resolution as the mean density is also lower (see Section~\ref{code}).


\section{Results}

\subsection{Galaxy morphologies}
\label{morphology}

\begin{figure*}

\includegraphics[width=6.5cm]{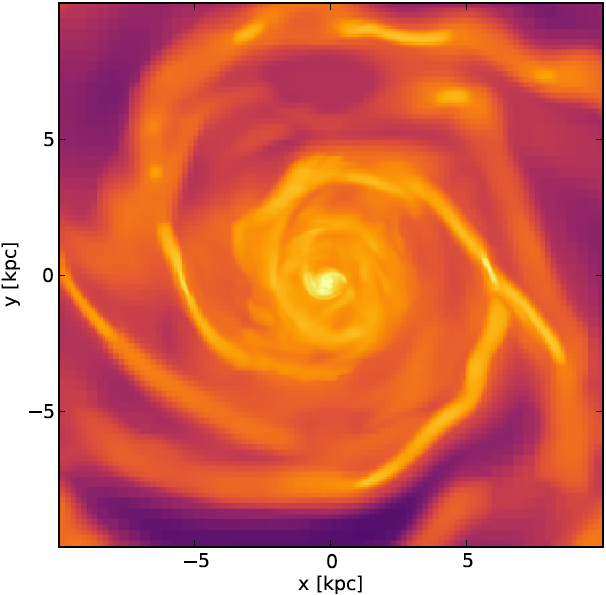}
\includegraphics[width=6.5cm]{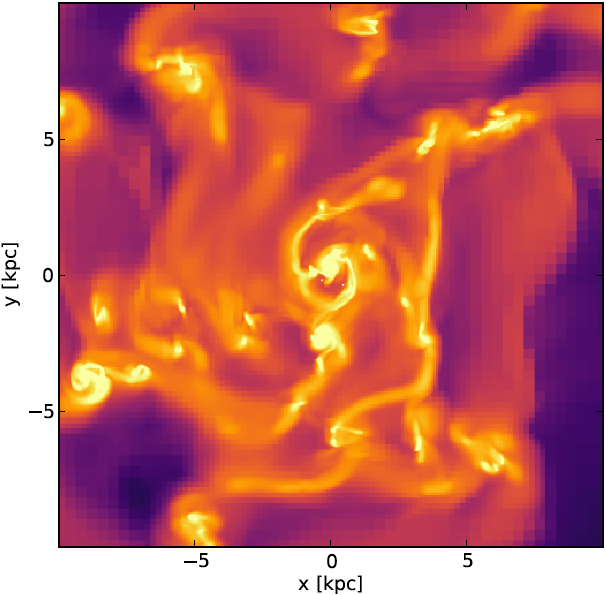}
\includegraphics[width=1.4cm]{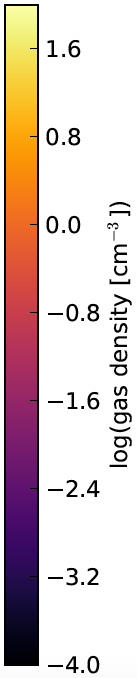}
\caption{Face-on mass-weighted mean gas density maps of the isolated runs: gp-iso is shown on the left panel and iso on the right panel. Both maps are obtained after an evolution during the time corresponding to the first pericentre for the orbit \#1 (490~Myr after the start of the simulation).\label{isomap}}
\end{figure*}

\begin{figure*}
\noindent \\
\includegraphics[width=4cm]{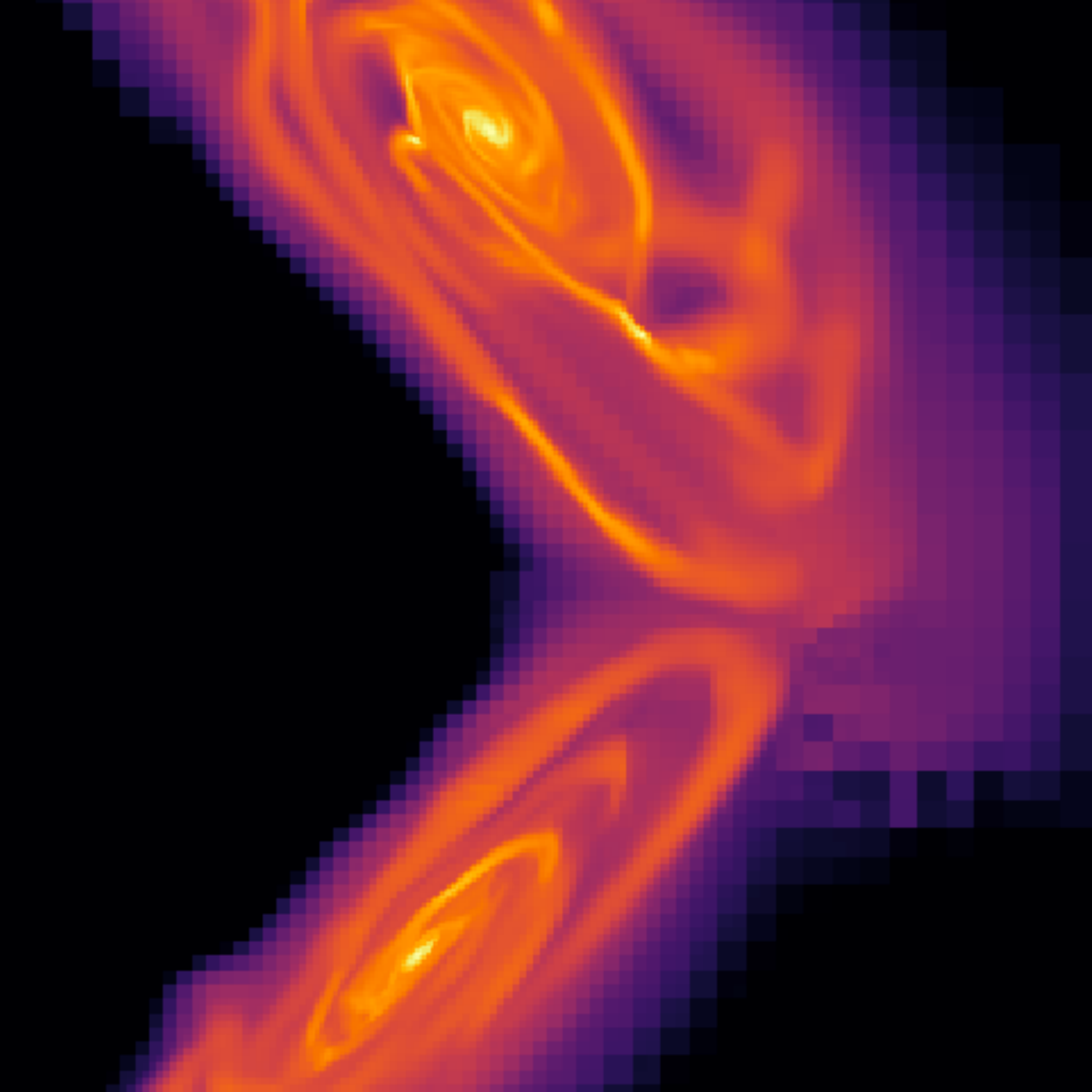}
\includegraphics[width=4cm]{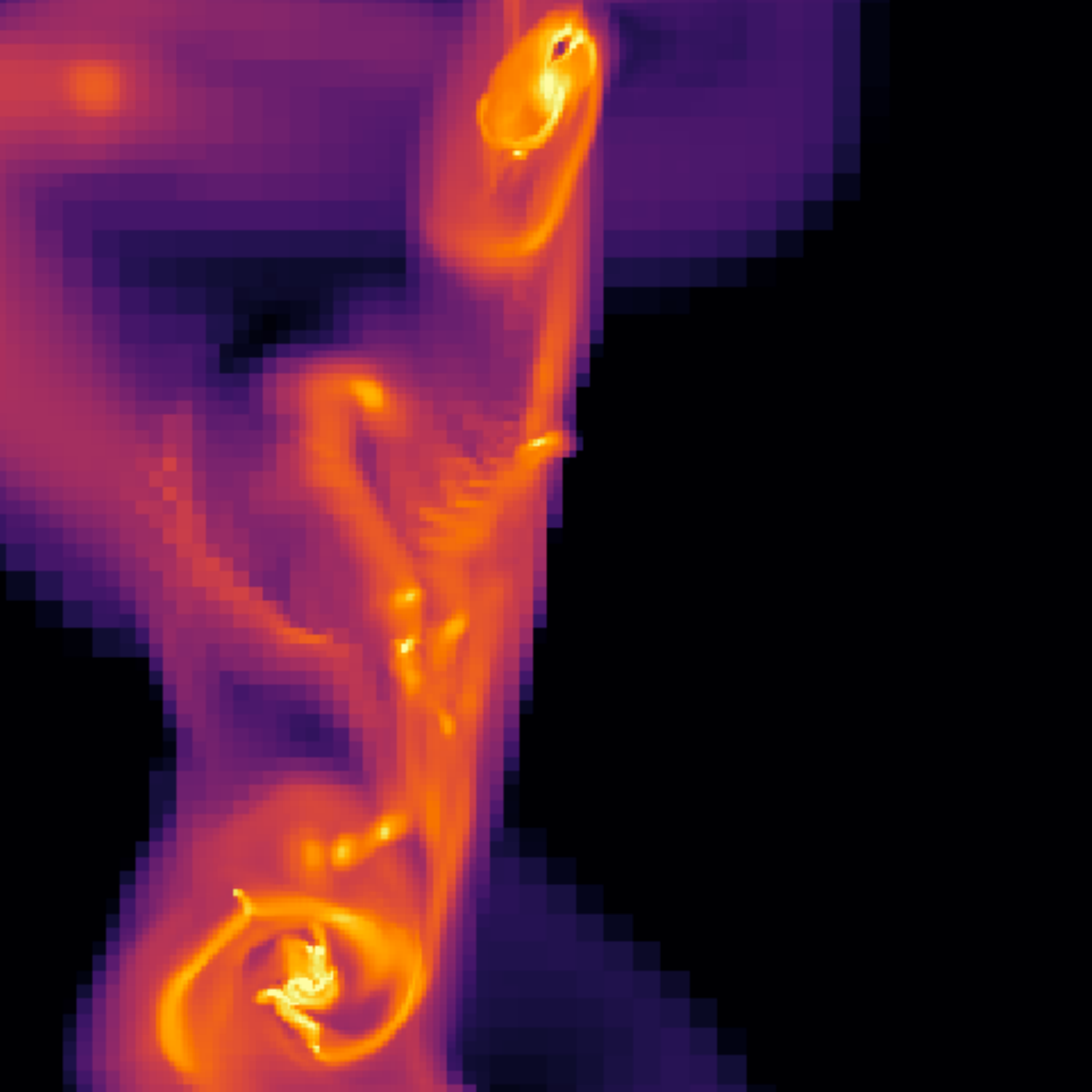}
\includegraphics[width=4cm]{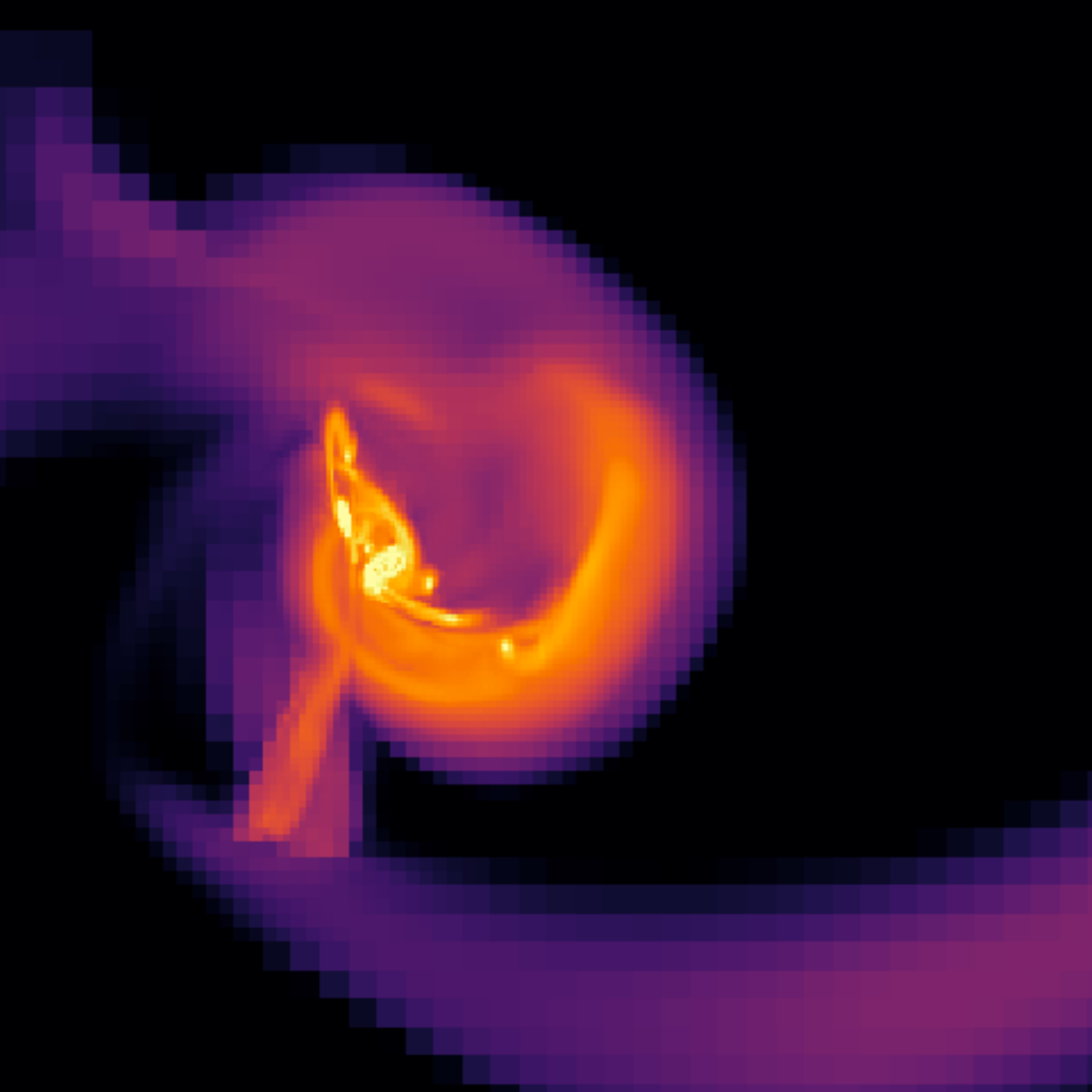}
\includegraphics[width=4cm]{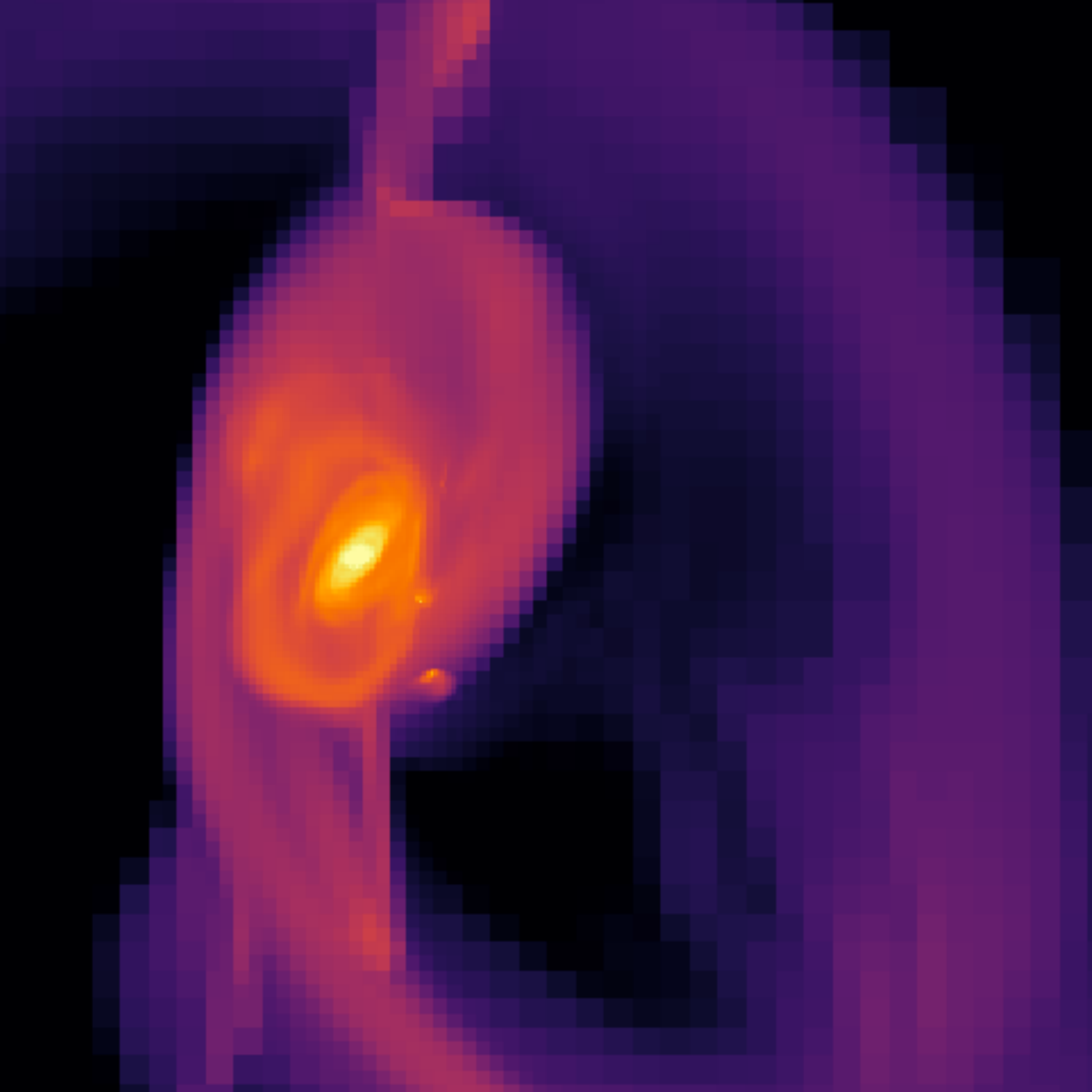}\\
\vspace{0.2 mm} 
 \includegraphics[width=4cm]{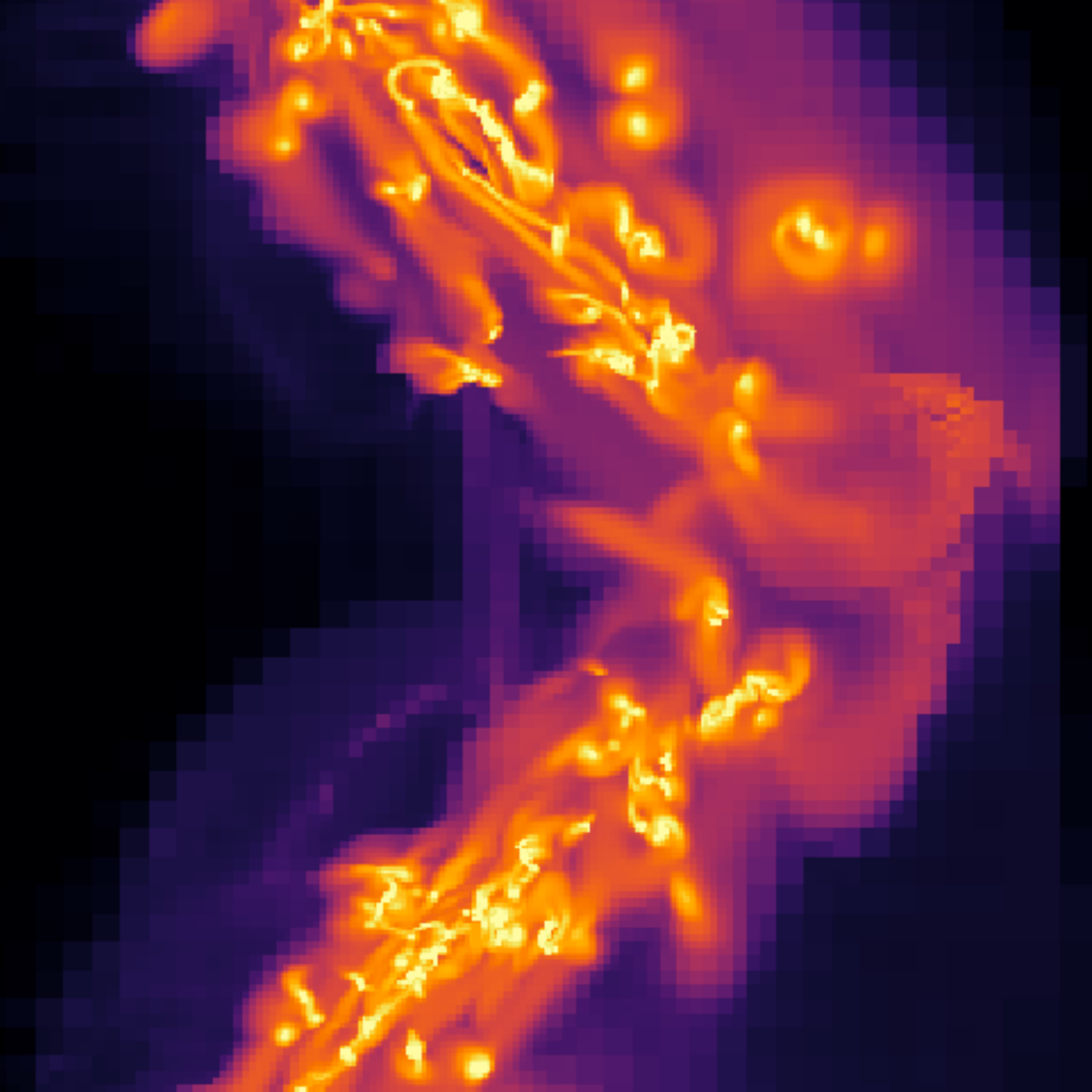}
\includegraphics[width=4cm]{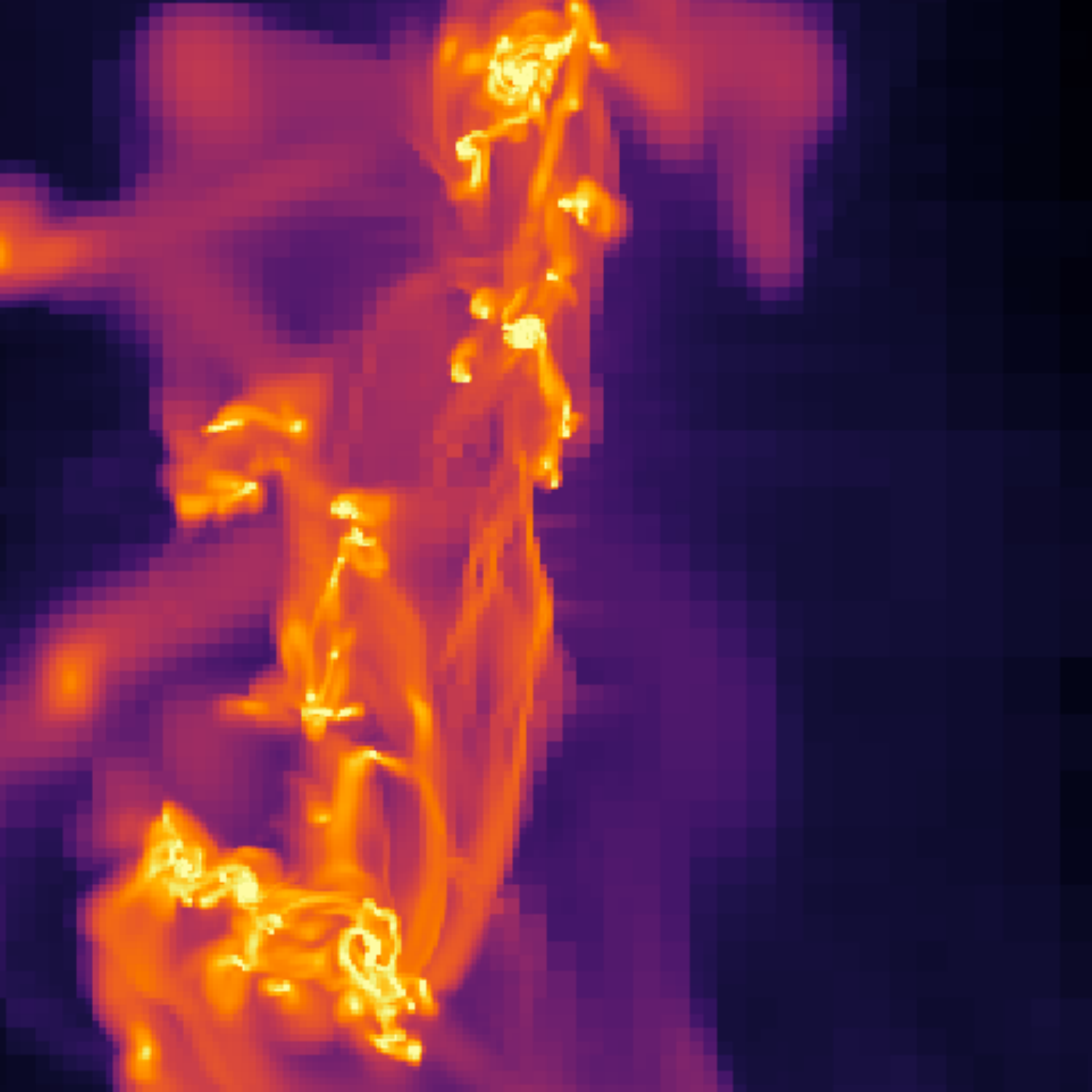}
\includegraphics[width=4cm]{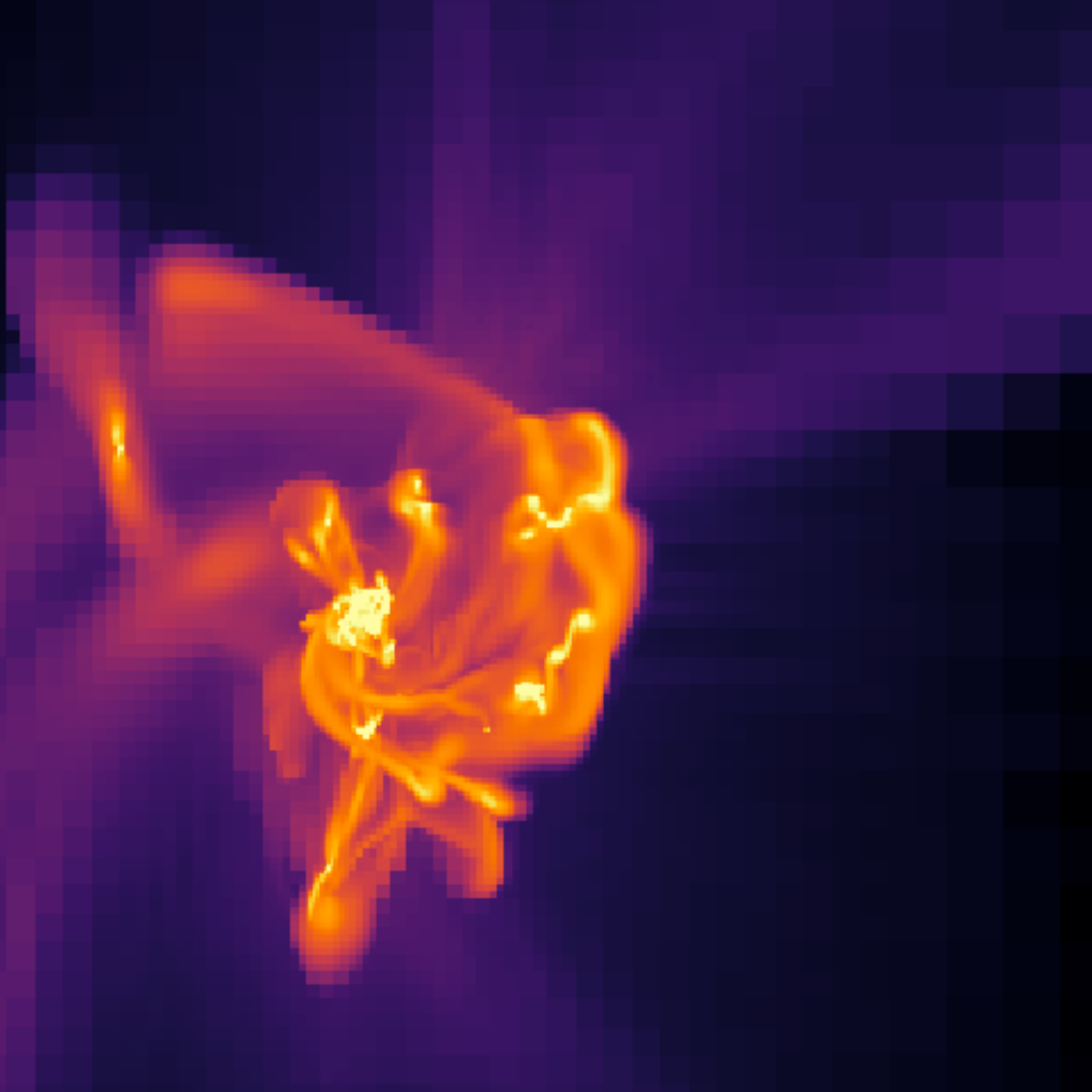}
\includegraphics[width=4cm]{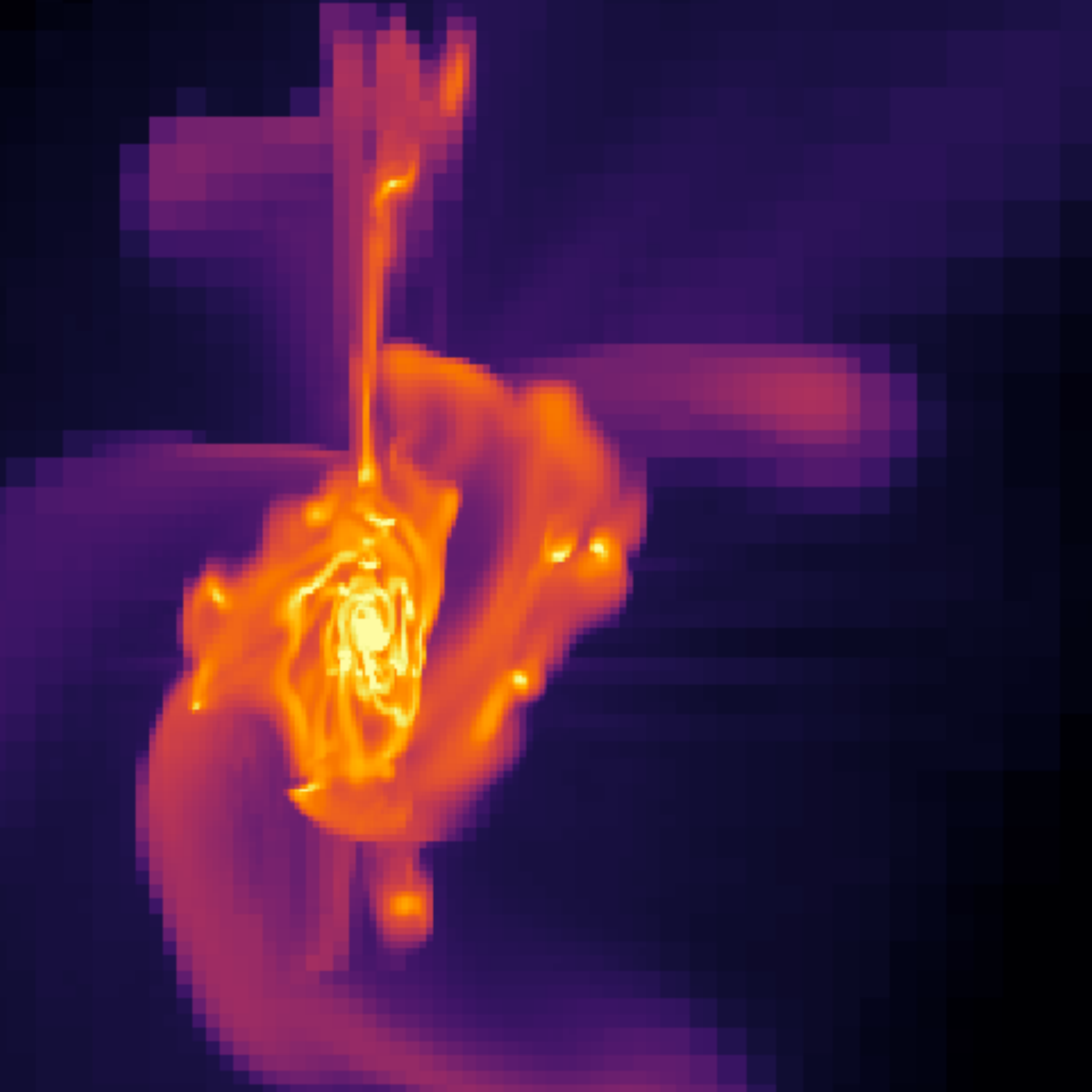}
\caption{Mass-weighted mean gas density map of the gas poor (top panel) and gas-rich (bottom panel) simulations for the direct-direct interaction orbit \#1 at different times from the first pericentre : -60, 105, 370 and 560 Myr from left to right. The maps span an area of 30~kpc $\times$ 30~kpc. The  colour scale is the same as in Fig.~\ref{isomap}.\label{comp}}
\end{figure*}

Fig.~\ref{isomap} shows a gas density map of the isolated run of both modelled disk galaxies. We see that while the low gas fraction disk shows several gaseous spiral arms, the high gas fraction disk is fragmented in several dense regions. These gaseous regions are observed to be long-lived\footnote{See movie at https://www.youtube.com/watch?v=ByV1g24eEjk}. They undergo merging and migration towards the centre with characteristic time of several 100 Myr. We will call them clumps thereafter.

In Fig.~\ref{comp} we show the gas density map of the merger in the low and high gas fraction case for the orbit \#1 (dd1 and gp-dd1), at different times of the interaction: a few 10~Myr before and after both the first pericentre passage and the coalescence. We see that the gas in the low gas fraction case shows the formation of a few clumps in the tidal bridge after the first pericentre passage. On the bottom row, the evolution of the gas density of the high gas fraction case does not show any significant increase of the number of clumps before and after the first pericentre passage. 

\begin{figure} 
\includegraphics[width=8.5cm]{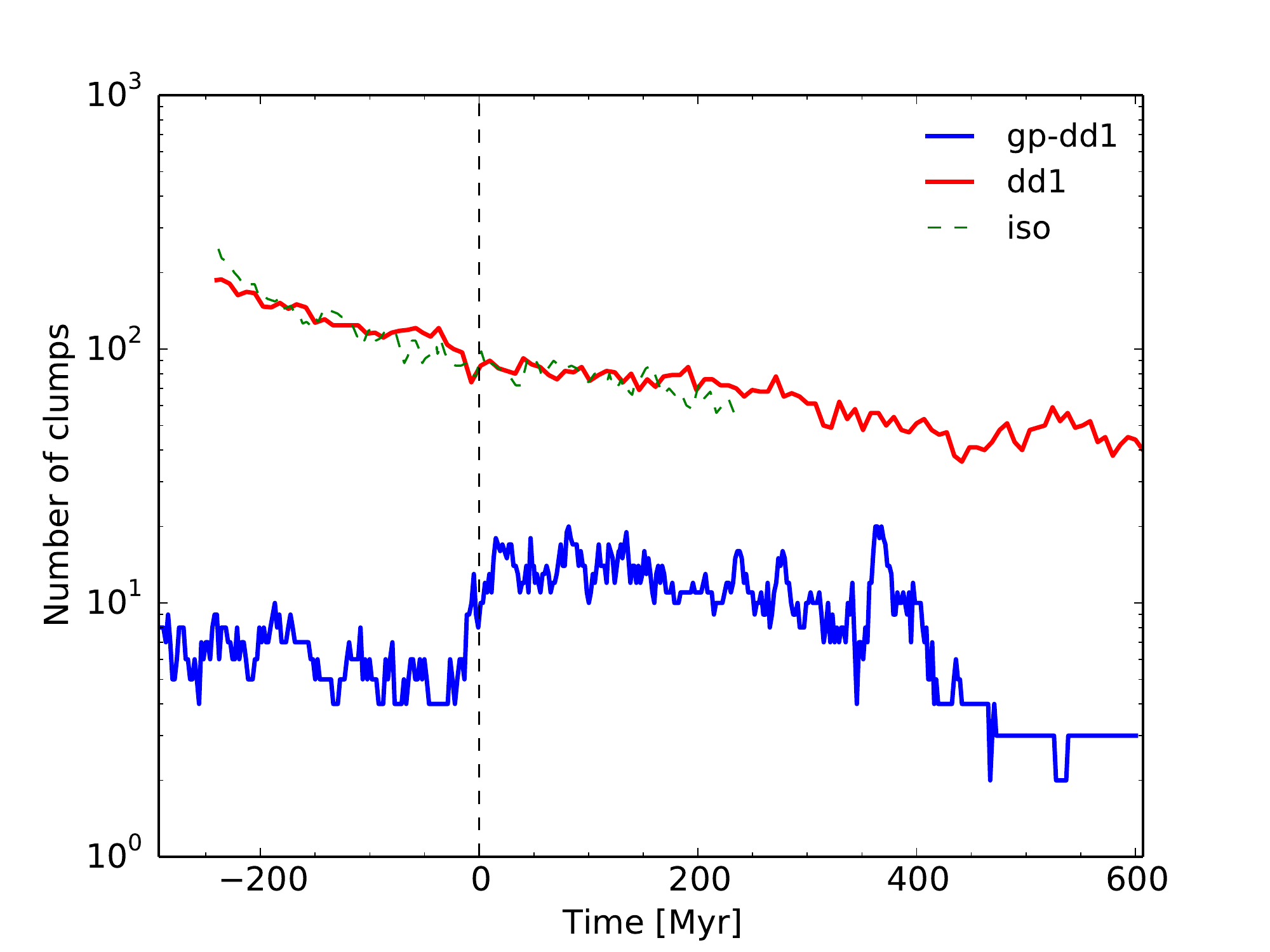}
\includegraphics[width=8.5cm]{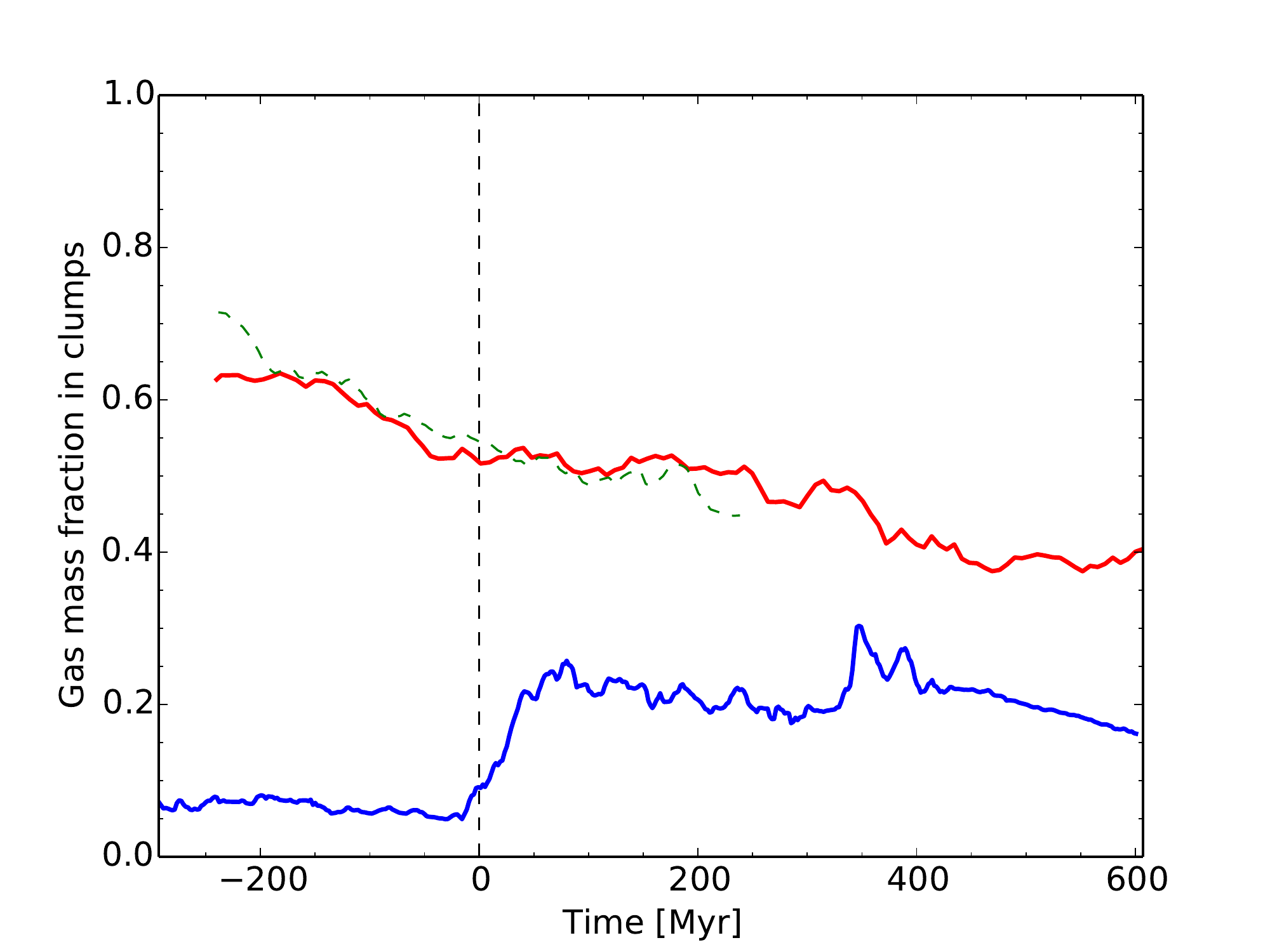}
\caption{{\it Top}: Number of detected clumps for the low (blue) and high (red) gas fraction for dd1 and gp-dd1. The isolated run for the high gas fraction case is also shown in dashed green. To ease the comparison, this number is multiplied by 2 for the isolated case. {\it Bottom}: Gas mass fraction in gas clumps.\label{clumps}}
\end{figure}

To have a more quantitative insight into the interaction-driven clumpiness of the disk, we use the friend-of-friend algorithm HOP \citep{Eisenstein98} to determine the number of clumps and the gas mass fraction embedded in them, during the interaction. Clumps are detected around peaks of gas density above 40~cm$^{-3}$. Sub-clumps are merged if the saddle density between them exceeds 4~cm$^{-3}$ and an outer boundary of  2~cm$^{-3}$ is set to define the density limit of the clumps. These parameters agree with a visual examination of the density maps and the results are not affected with respect to small changes of these parameters.

In Fig.~\ref{clumps} we show the evolution of the number of clumps, and the gas fraction inside them. In the low gas fraction case the gas fraction enclosed in clumps goes from 5 to 25\% at the first pericentre passage, and the number of clumps increases by a factor 4, from about 4 to 16.  These numbers stay relatively constant along the interaction. This enhanced number of clumps in low redshift galaxy collisions has been already noted by \citet{DiMatteo08}. In the high gas fraction case, the gas mass fraction in clumps shows instead a slow but steady decrease of both the number of clumps and gas mass fraction enclosed. The evolution of the number of clumps and masses of the high gas fraction case are therefore not influenced by the interaction as we can see by comparing the isolated (iso) and the interacting cases (dd1). In particular, no change is observed at the first pericentre passage.


\subsection{Evolution of the density PDF}
\label{sectionPDF}

\begin{figure*} 
\includegraphics[width=8.5cm]{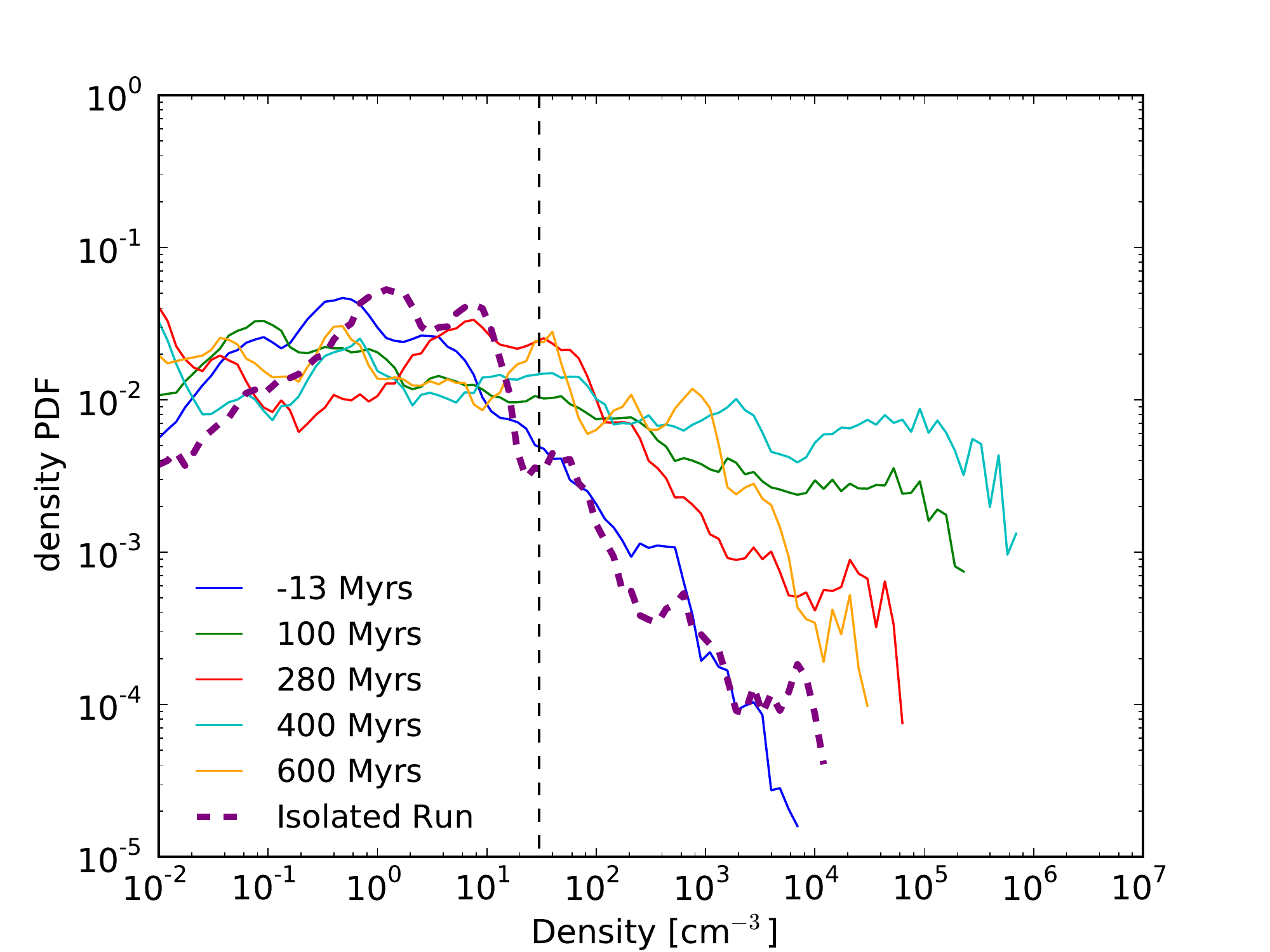}
\includegraphics[width=8.5cm]{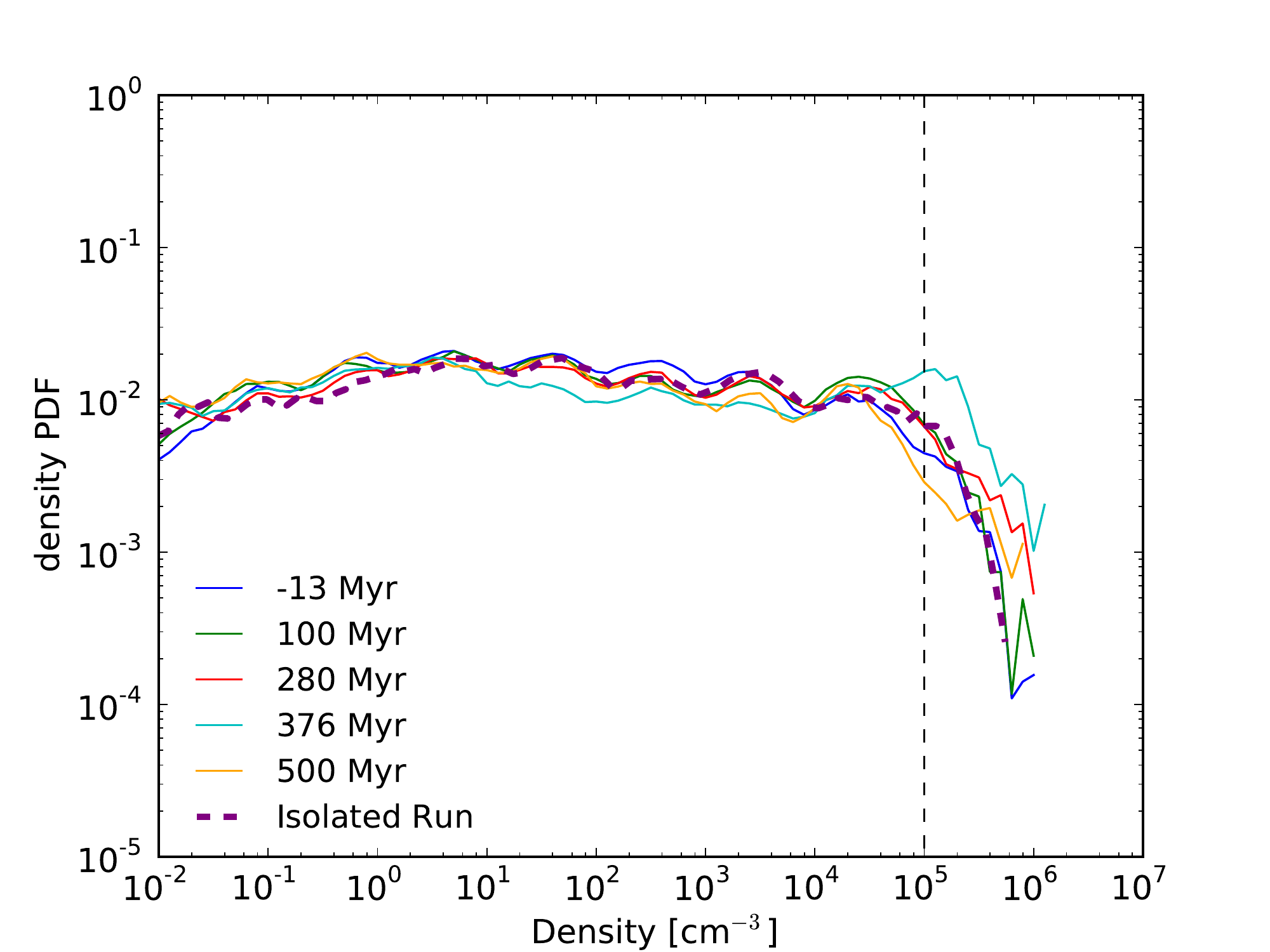}
\caption{Normalized mass-weighted probability distribution
function of the gas density, at several epochs  ($t=0$ corresponds to the first pericentre). These times correspond to before the first pericentre, shortly after the first pericentre, before the coalescence, peak of SFR during coalescence, long after the coalescence, respectively. Curves are shown for the direct-direct encounter on orbit\#1 for the low (left) and high (right) gas fraction cases in solid lines. The thick purple lines show the corresponding isolated cases. The black dashed line indicates the density threshold for star formation as defined in Section ~\ref{code}.\label{PDF}}
\end{figure*}

In the low gas fraction, this interaction-induced clumpiness condensates the gas in localized over densities. To quantify its effect on the gas density distribution, we look at the evolution of the mass-weighted density probability density function (PDF). In Fig.~\ref{PDF}, we see the evolution of the shape of the gas PDF for both the low and high gas fraction cases. In the low gas fraction case, the PDF extends to higher densities during the interaction. We also see that the PDF of the high-gas fraction case remains almost identical during the interaction and does not significantly vary from the shape of the isolated galaxy PDF, except at the coalescence ($t = 376$ Myr) where there is an increase of the mass with densities above 3 $\times 10^{4}$~cm$^{-3}$.

The evolution of the shape of the PDF is of particular interest, as high density gas fuels star formation. Indeed, in the low gas fraction case, the gas mass above the density threshold goes from 4\% to 20\% after the first pericentre passage and 40\% at the coalescence, while it stays constant, at 2-3\%, in the high-gas fraction case.

\subsection{Merger-induced star formation}

\subsubsection{Star formation histories}
\label{sb}

\begin{figure*}
\includegraphics[width=8.5cm]{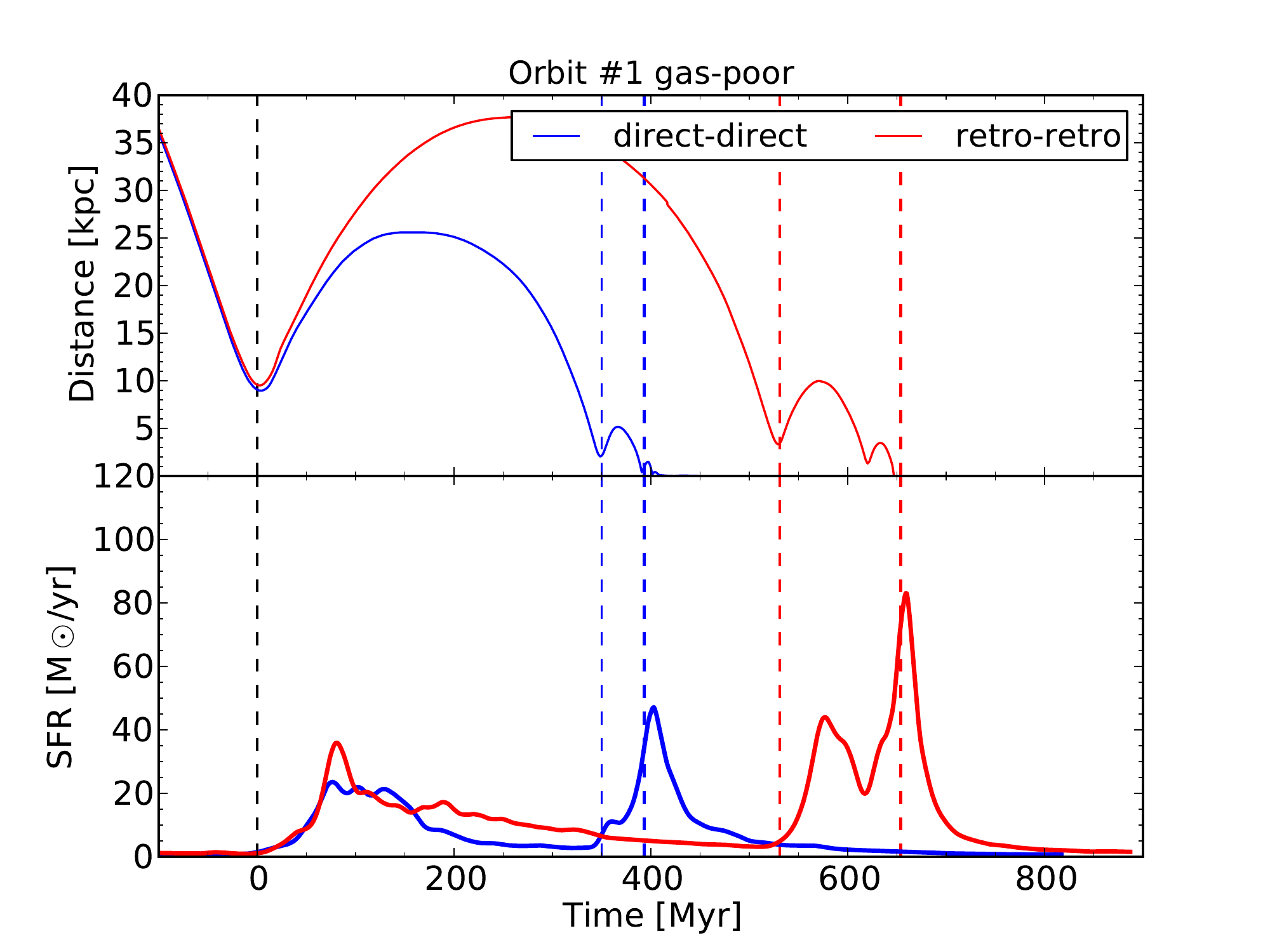}
\includegraphics[width=8.5cm]{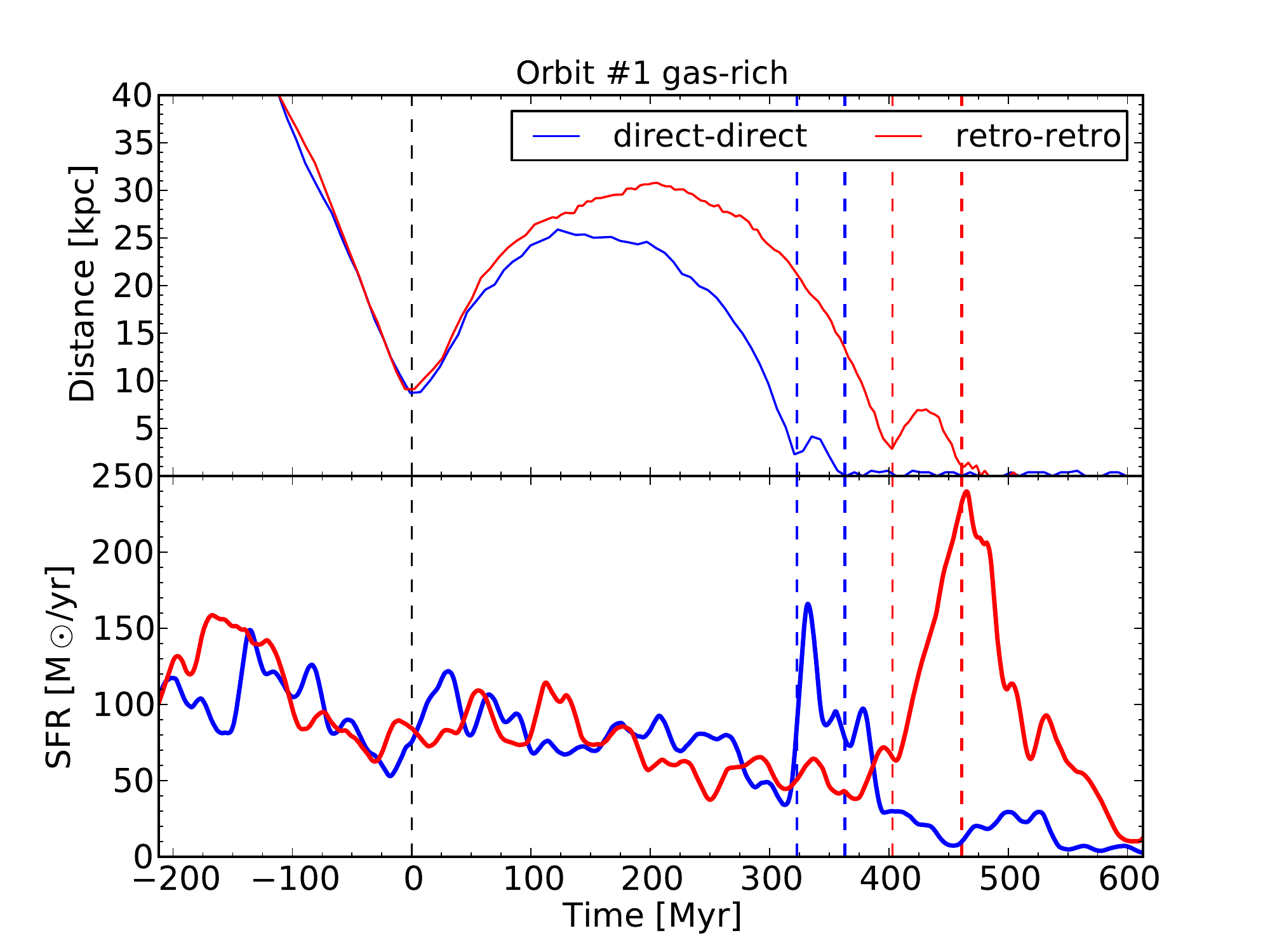}
\caption{Star formation rate during the simulations run on orbit \#1, with the distance between the galaxies plotted above. The direct-direct and retro-retro simulations are shown respectively in blue and red. The low gas fraction case is shown on the left panel and the high-gas fraction case on the right panel. The dashed lines correspond to the time of the first pericentre passage (in black) and the respective times of second pericentre passage and coalescence (in the  colour corresponding to the simulation). \label{compSFR}}
\end{figure*}

\begin{figure*}
\includegraphics[width=8.5cm]{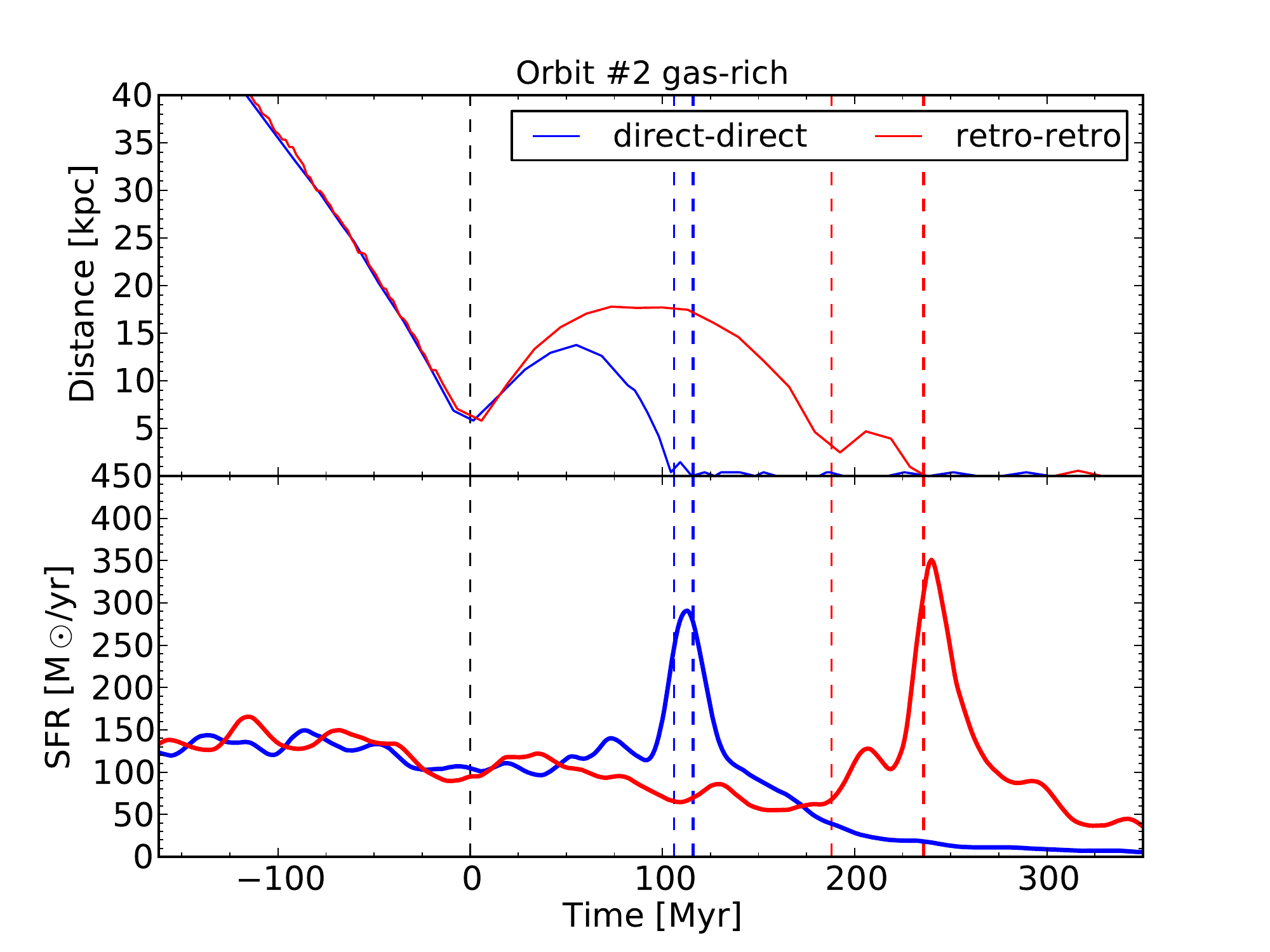}
\includegraphics[width=8.5cm]{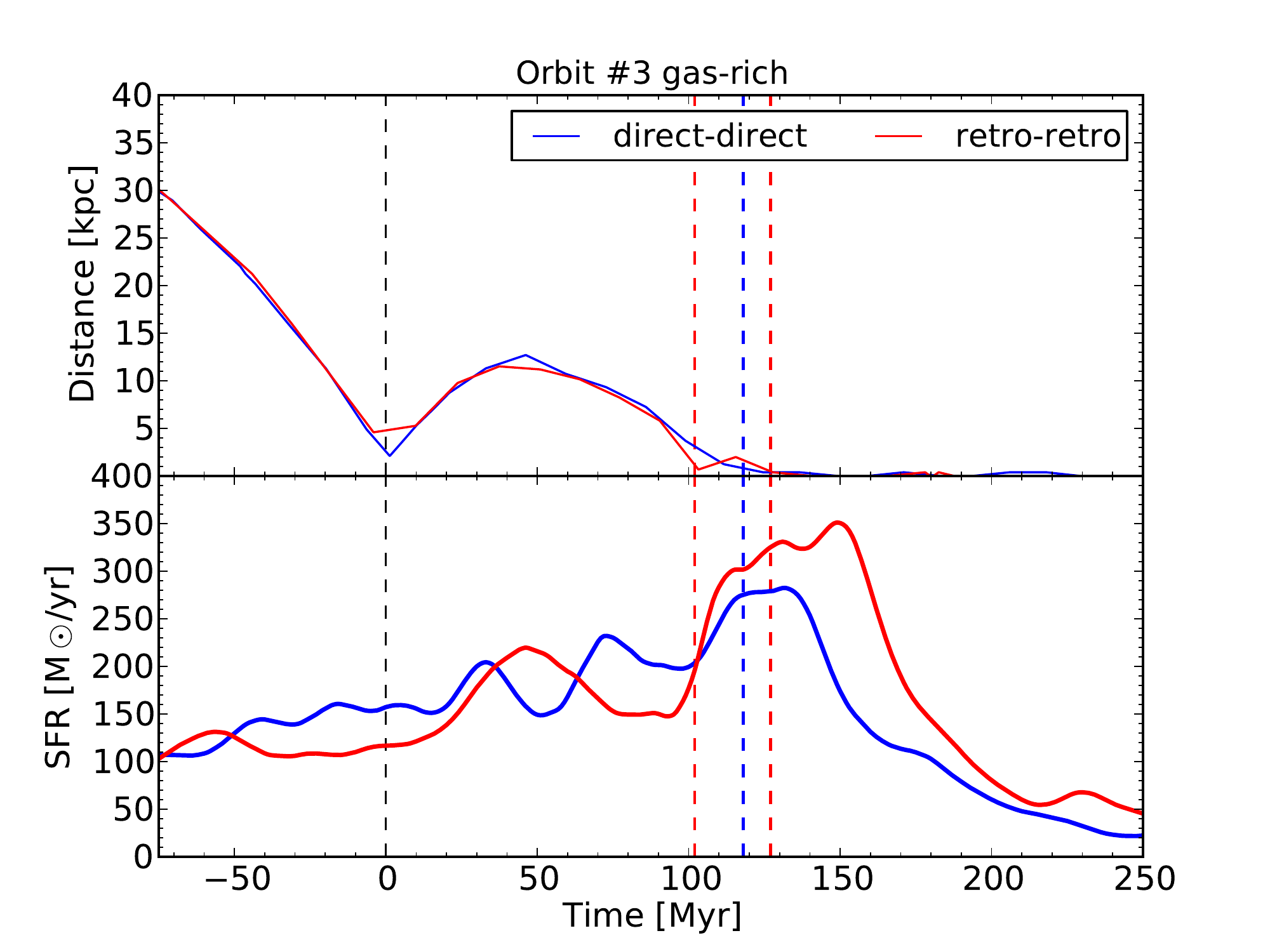}
\includegraphics[width=8.5cm]{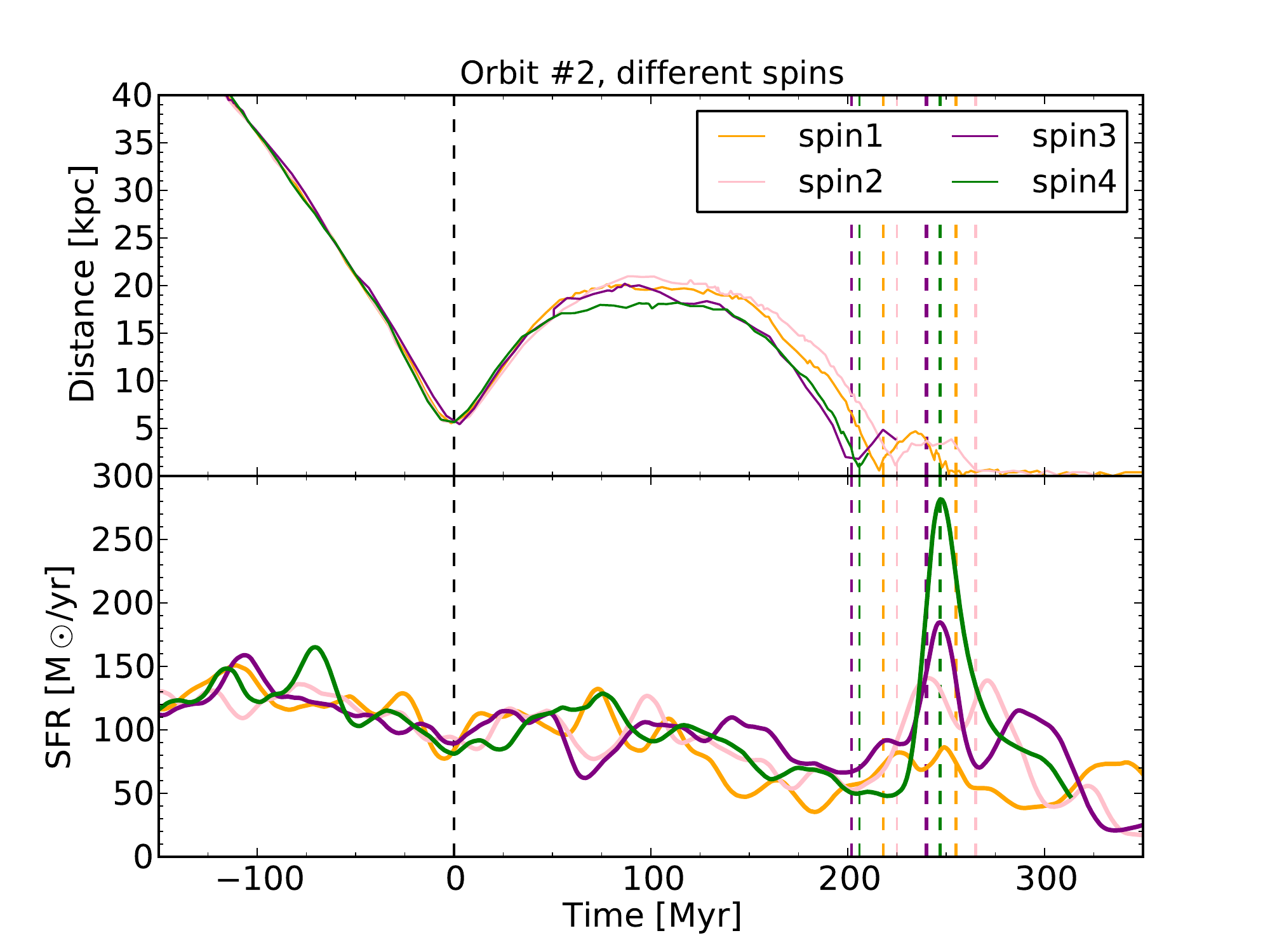}
\caption{Same as Fig.~\ref{compSFR} but for the orbit \#2 and orbit \#3 (top panels). The bottom panel show the star formation rate for different spin orientations. The vertical dashed lines are the same as in Fig.~\ref{compSFR}. \label{SFR}}
\end{figure*}

The comparison between low and high gas fraction star formation histories during the interaction is shown in Fig.~\ref{compSFR}. In the low gas fraction case, the SFR is about 1-2~M$_{\odot}$/yr before the interaction and increases to $\sim$ 30~M$_{\odot}$/yr and  more than 50~M$_{\odot}$/yr after the first pericentre passage. The SFR then lowers back to a few  M$_{\odot}$/yr before increasing again to $\sim$ 40 to 60~M$_{\odot}$/yr  during the final coalescence.

The high gas fraction case shows a very different history. The SFR is initially relatively high, $\simeq$ 120 M$_{\odot}$/yr for the galaxy pair, and does not increase significantly at the first pericentre passage. The SFR is only enhanced at the coalescence\footnote{In both gas fraction cases, the retrograde-retrograde simulation coalescence happens later than in the direct-direct simulation. This is expected, because they do not form tidal tails, which are efficient in driving angular momentum out.}, and only by a factor at most 5. Increasing only the gas fraction thus appears to significantly lower the boost of SFR. 

To ensure that this results does not depend on a particularity of the chosen orbit or orientation we also look at the SFR of the two other tested orbits and orientations, which are displayed in Fig~\ref{SFR}. We see that they follow the behavior of orbit \#1: they all show almost no increase of star formation at the first pericentre passage, and a mild burst at coalescence. 


\subsubsection{The starburst sequence}

\begin{figure*} 
\includegraphics[width=8.7cm]{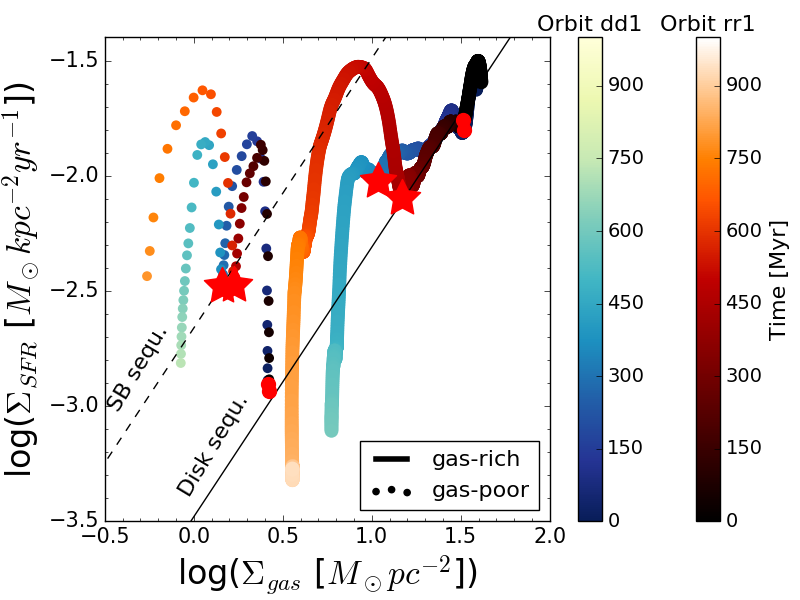}
\includegraphics[width=8.7cm]{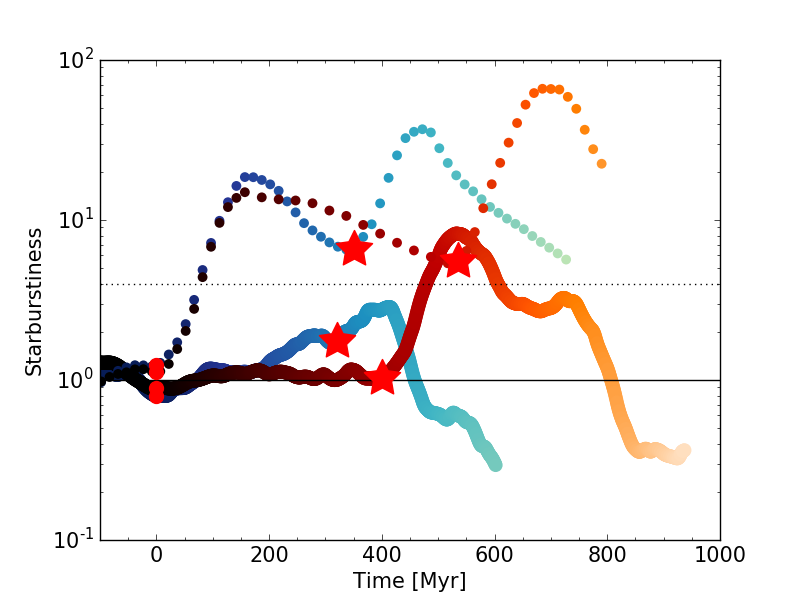}
\caption{{\it Top panel: } Evolution of the simulations dd1 (shades of blue) and rr1 (shades of red) on the Schmidt-Kennicutt diagram. The solid lines show the high gas fraction cases, and the dotted lines the low gas fraction cases. The red dots and stars show respectively the time of the first and second pericentre. The solid and dashed black lines are respectively the disk and starburst sequences from \citet{Genzel10}. The curve is smoothed using a constant kernel on the previous 100~Myr, to resemble the smoothing of the measurements of the SFR using IR luminosity \citep{Kennicutt98}. {\it Bottom panel:} Evolution of the starburstiness, as defined in the text, during the interaction. The  colours and labels are the same as in the upper panel. The black solid line indicates a starburstiness of unity, i.e a SFR equivalent to the disk sequence for a given , $\Sigma _{\mathrm{gas}}$, and the dotted line shows a starburstiness of 4, which we define as the lower limit to be considered as a starburst galaxy\label{ks}.} 
\end{figure*}

Our motivation to run these simulations resides in the fact that a smaller than expected number of high-redshift galaxies are found on the starburst sequence of both the $M_{\star}$--SFR and $\Sigma _{\mathrm{gas}} $--$ \Sigma _{\mathrm{SFR}}$ diagrams, the latter being also known as the Schmidt-Kennicutt diagram.

In Fig.~\ref{ks} we can see the evolution of our simulated galaxies on the Schmidt-Kennicutt diagram. The computation is done in a box of 60~kpc$\times$~60~kpc~$\times$~60~kpc and we rescale the curves so that the pre-merger disks lie on the disk sequence. To better quantify the starbursting behavior of our simulations we define the starburstiness parameter as the measured $\Sigma _{\mathrm{SFR}}$ over the value of $\Sigma _{\mathrm{SFR}}$ corresponding the the disk sequence of \citet{Genzel10} for the measured $\Sigma _{\mathrm{gas}}$. We define as starbursting a galaxy with a starburstiness exceeding ,4 meaning that the galaxy stands more than 0.6 dex above the disk sequence, which is a common definition for the starburst sequence \citep[see e.g.][]{Schreiber15}.

We see that the galaxies follow the disk sequence until the collision which shifts their loci towards the starburst sequence. The low gas fraction systems reach the starburst sequence already at the first pericentre passage and their starburstiness stays above 4 all along the starburst, that is for more than 700~Myr for the considered orbit. The high gas fraction pairs hardly reach the starburst sequence: the gas-rich direct-direct starburstiness always stays below 4 and is thus never considered as a starburst galaxy. The burst at the coalescence of the gas-rich retrograde-retrograde encounter makes it reach the starburst sequence of the Schmidt-Kennicutt diagram, but only for $\sim$ 100~Myr.\\

In the following we analyse the processes that influence the ability of high fraction gas merger to trigger starbursts, as compared to low gas fraction ones.


\section{What causes this weak enhancement of  star formation?}


\subsection{Weak increase of the central gas inflows}

\begin{figure}
\includegraphics[width=9cm]{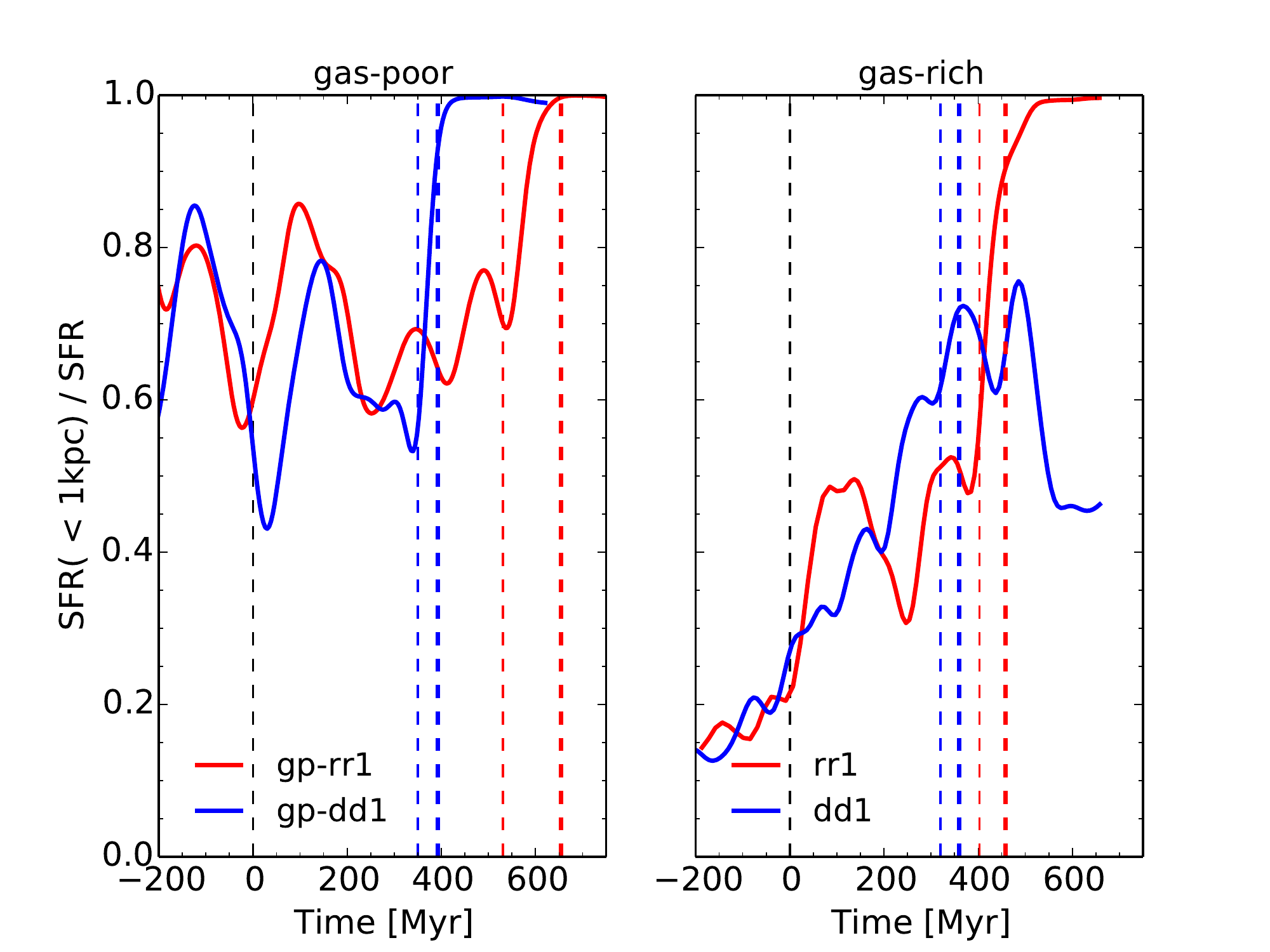}
\caption{Evolution of the fraction of SFR located inside the central kpc of both galaxies for the encounters on orbit \#1 in the low gas fraction case (left) and high gas fraction cases (right). The vertical dashed lines are the sane as in Fig.~\ref{compSFR}.\label{insfr}}
\end{figure}

\begin{figure*}
\includegraphics[width=8cm]{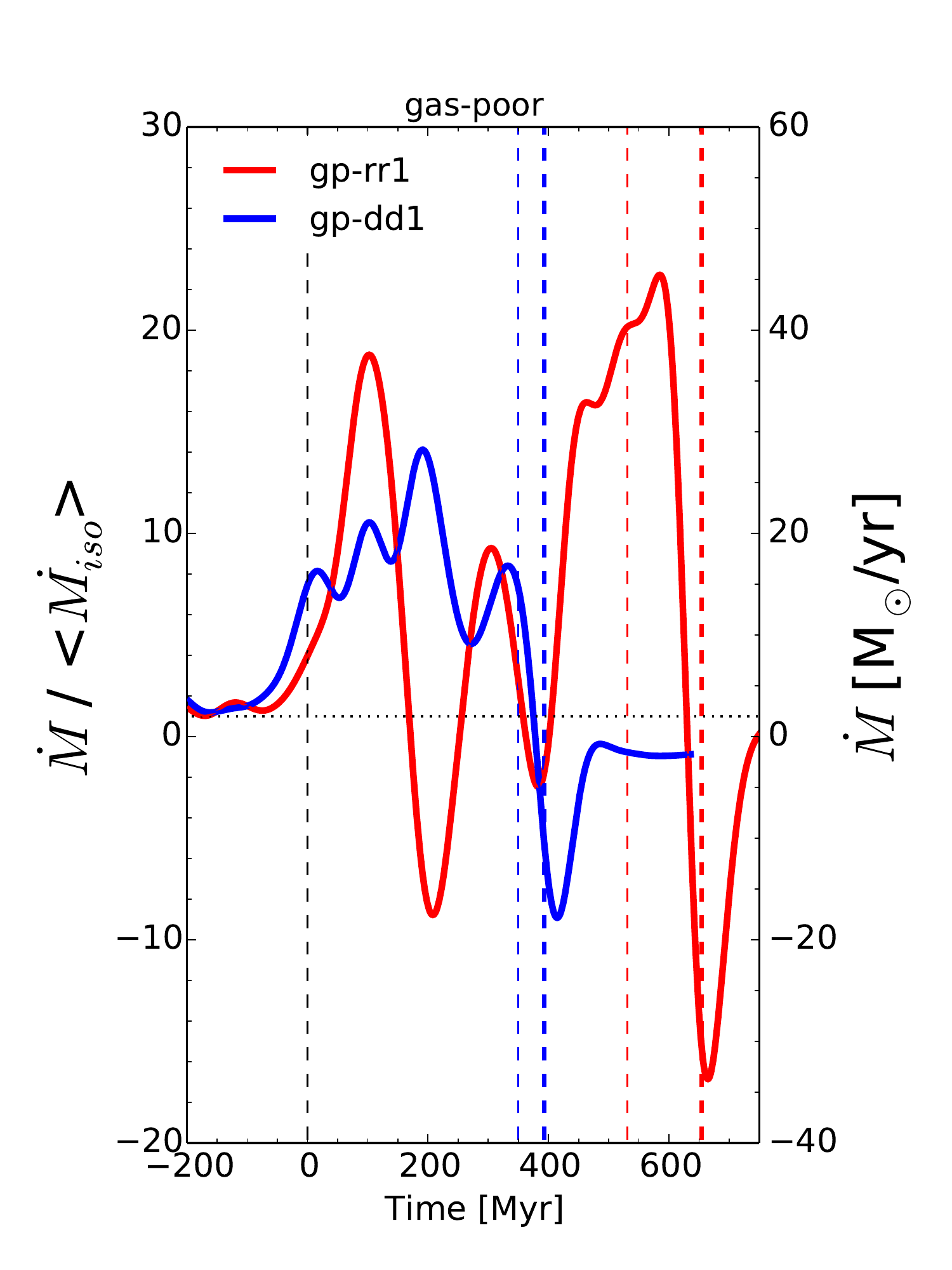}
\includegraphics[width=8cm]{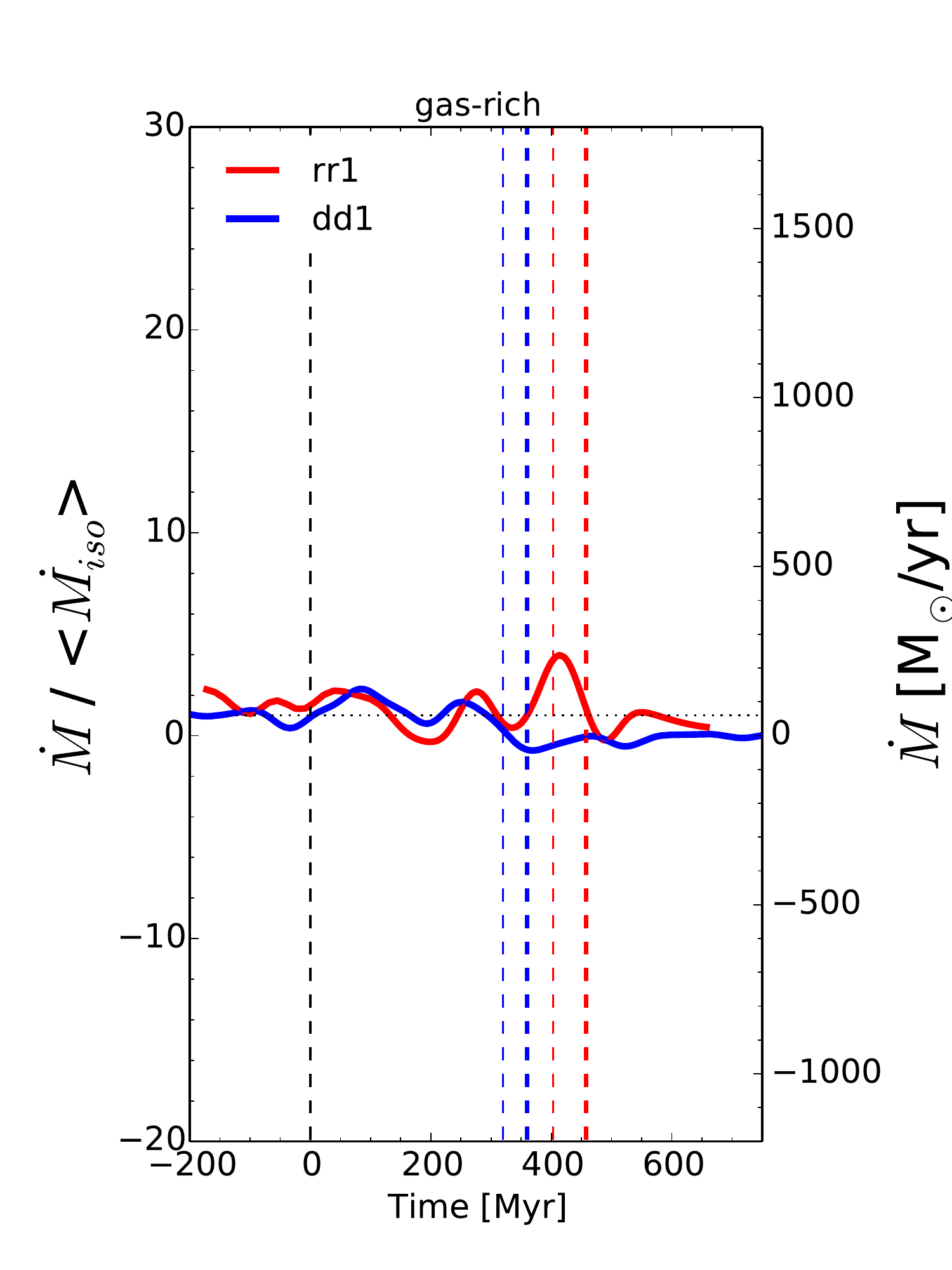}
\caption{Evolution of the inflow of baryonic mass fraction in the central kpc of both galaxies for the encounters on orbit \#1 in the low gas fraction case (left) and high gas fraction case (right). The left axis shows the ratio of the inflow to the mean inflow rate of the isolated run, and the same scale is used for both plots to emphasize the relative difference to the isolated case. The right axis shows the actual value of the inflow rate. The curves are smoothed using a gaussian kernel of root mean square width 30 Myr for the sake of clarity. The vertical dashed lines are the same as in Fig.~\ref{compSFR}.\label{inflows}}
\end{figure*}

Observed starbursting galaxies show a prominent nuclear starburst \citep{Sanders96}, with an important concentration of gas in the central kpc of the galaxies. Fig.~\ref{insfr} shows the fraction of SFR located in the central kpc of the galaxies. We see that at the coalescence, the peak of star formation is almost entirely located in this central kpc\footnote{with the exception of dd1 simulation: in Fig.~\ref{comp} one can actually see in the bottom-right corner that this particular simulation forms an irregular fragmented structure, surrounded by dense, star-forming gas clumps more than 1~kpc away from the center, which explains why star formation is not confined in the central kpc only.}. 

This centrally concentrated star formation is fueled by gas inflows towards the center. Interaction-driven gravitational torques are indeed expected to drive large amount of gas towards the nuclei of both galaxies \citep{Barnes91, Mihos96}. In Fig~\ref{inflows} we show the inflows of baryonic mass in the central kpc of our galaxies. For the low gas fraction case, pre-merger disks have central mass inflows of $\simeq$ 2-3~M$_{\odot}$/yr. Between the first and second pericentre the disks are perturbed and the inflows have a higher mean value of about 20~M$_{\odot}$/yr, with some important variation over time. At coalescence, the inflows reach a peak at 20~M$_{\odot}$/yr for the dd1 and 50~M$_{\odot}$/yr for the rr1 simulation, which are both at least ten times higher than the initial gas inflows in the isolated case.

For the high gas fraction case, the central inflows are stronger from the beginning, $\simeq$ 30 to 50~M$_{\odot}$/yr. As in the low gas fraction case, a strong peak in inflows is seen for the rr1 simulation, at 250~M$_{\odot}$/yr. The relative increase compared to the pre-merger case is less than 5, much less than for the low gas fraction case. The resulting increase in SFR is also comparatively less important than for the low gas fraction case.

Central gas inflows are already strong in the pre-merger high gas fraction disk because of the violent disc instability (VDI): the high turbulence of the gas is in part fueled by inwards the migration of gas \citep{Dekel09b}. Another physical explanation for the higher increase of the central inflows in low gas fraction discs is the presence of a more massive stellar component which creates a tidally-induced bar or spiral arms which exert strong torques on the gas, and which adds to the effect of the gravitational torques originating from the companion. The stellar component being less populated in the high gas fraction case, this process is less efficient, as already noted by \citet{Hopkins09}.

The high gas fraction, which leads to a high turbulence and the formation of clumps, also drives strong gas inflows in the isolated disks. As a result, interaction-induced gas inflows are less important with respect to pre-merger inflows for high gas fraction disks than for low gas fraction disks. This leads to a lower increase of SFR due to central gas inflows.

\subsection{Mild enhancement of gas turbulence}
\label{turb}

\begin{figure} 
\includegraphics[width=9cm]{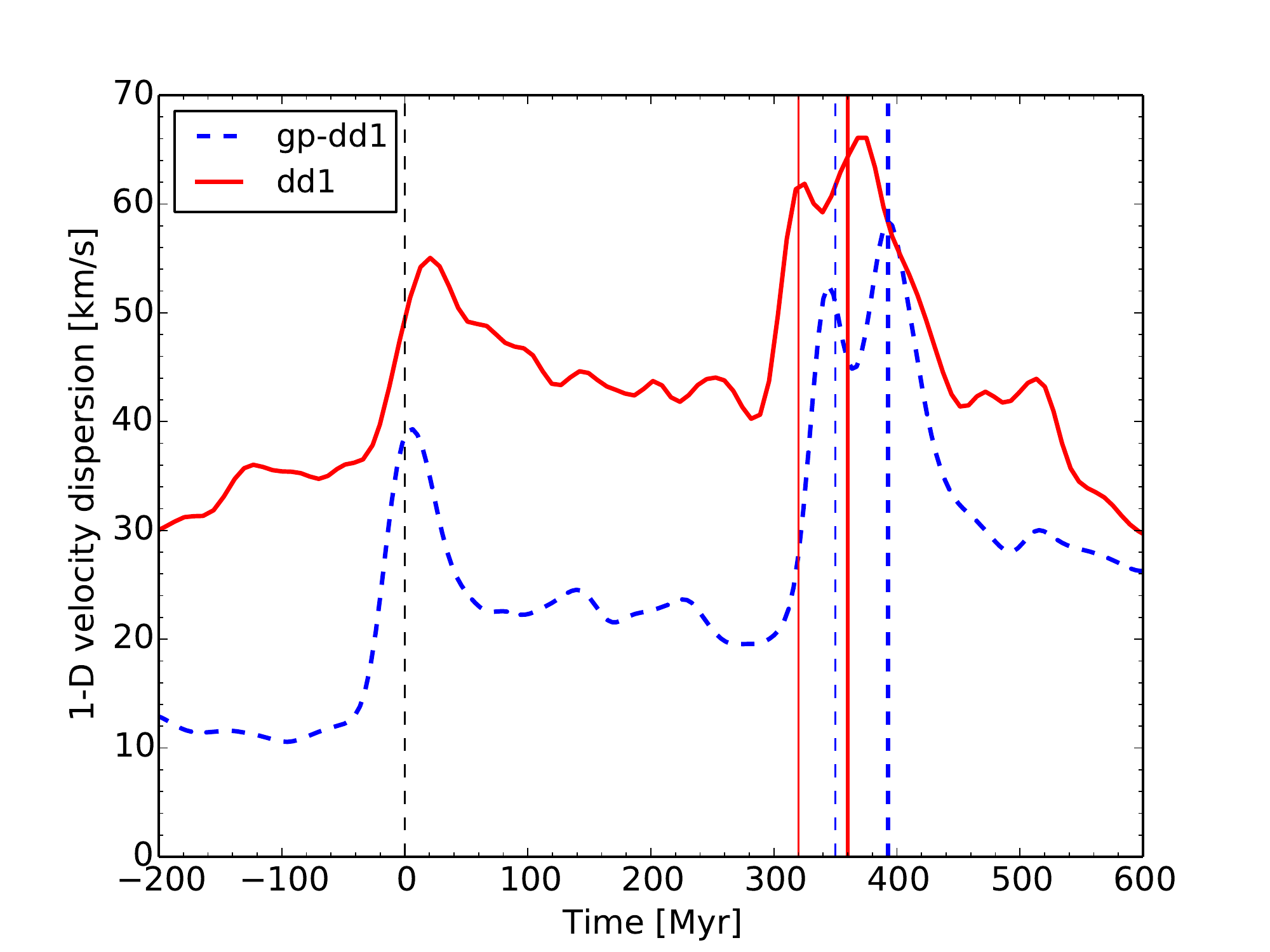}
\caption{ Evolution of the gas velocity dispersion of the direct-direct simulations on orbit \#1. The low gas fraction is depicted by the blue dashed curve and the high gas fraction by the solid red line. The vertical dashed lines are the same as in Fig.~\ref{compSFR}.\label{disp}}
\end{figure}

Our current understanding of extended starburst in low gas fraction galaxy interactions is that the increase of the SFR is caused by an increase of gas turbulence.

We compute the 1-dimensional velocity dispersion of the gas, along the line of sight of the simulations at the scale of 100~pc and plot its evolution in Fig.~\ref{disp}. We see that the pre-interaction velocity dispersion is much higher in the high gas fraction case, $\sigma \simeq$~35~km/s, than in the low gas fraction case, $\sigma \simeq$~10~km/s. This agrees with observations \citep{Forster09} and is due to the gas phase which is more gravitationally unstable with high gas fraction. The interaction brings the velocity dispersion of the low gas fraction case to more than 40~km/s, in agreement with both observations \citep{Elmegreen95, Elmegreen16} and previous simulations \citep[][ R14]{Bournaud08}. In the high gas fraction case  the velocity dispersion goes up to around 60 km/s only. The relative increase in turbulence is much higher in the low gas fraction case, a factor 5, to be compared to the high gas fraction case, which shows an increase of less than a factor 2.

If we suppose, for convenience, that the gravitational specific energy is equally transferred between the gas, stars and dark matter during the interaction, we can write, for a timescale much shorter than that of the dissipation of the turbulence at the scale of clumps (i.e 
$\ll$ 10Myr):

\begin{equation}
\frac{3}{2} M _{\mathrm{gas}} \sigma_{f}^{2} = \frac{3}{2} M _{\mathrm{gas}} \sigma_{i}^{2} + M _{\mathrm{gas}}~f\Delta \phi
\end{equation}
 where $\sigma_{i}$ and $\sigma_{f}$ are the initial and final 1-dimensional velocity dispersions, $\Delta \phi$ is the difference in gravitational potential, and $f$ the fraction of this potential energy which is transferred to the turbulent motion during the interaction, and which we suppose to be a constant. After the above assumptions, $M _{\mathrm{gas}}$ disappears on both sides of the equation, which therefore applies for both gas fraction regimes. It follows that, for the same $f \Delta \phi$:

\begin{equation}
\sigma_{f} = \sqrt{ \sigma_{i}^{2} + \frac{2}{3}~f \Delta \phi}
\end{equation}
for both the low and high gas fraction case. Hence, if $f \Delta \phi$ brings $\sigma$ from 10 km/s to 40 km/s in the low gas fraction case, it will increase $\sigma$ from 40 km/s to $\simeq$ 55 km/s in the high gas fraction case, which is approximately what we measure in our simulations. It is interesting to note that the required value of $f \Delta \phi$ to increase $\sigma$ from 10 to 40 km/s in our calculation happens to be less than the specific gravitational energy liberated in 10~Myr (time-scale for dissipation of turbulence) for the masses considered here, when the distance between the two galaxies is smaller than 25~kpc, i.e $\sim$ 50~Myr before the first pericentre for orbit \#1/. This is the correct order of magnitude for the onset of the rise of the velocity dispersion in our simulations  (see Fig.~\ref{disp}). 

This heuristic calculation shows the difficulty of increasing the velocity dispersion while it is already high, which leads to a mild increase in gas turbulence from interactions in high gas fraction disks. 
 

\subsection{Absence of interation-induced tidal compression}

\begin{figure*}
\includegraphics[width=8.5cm]{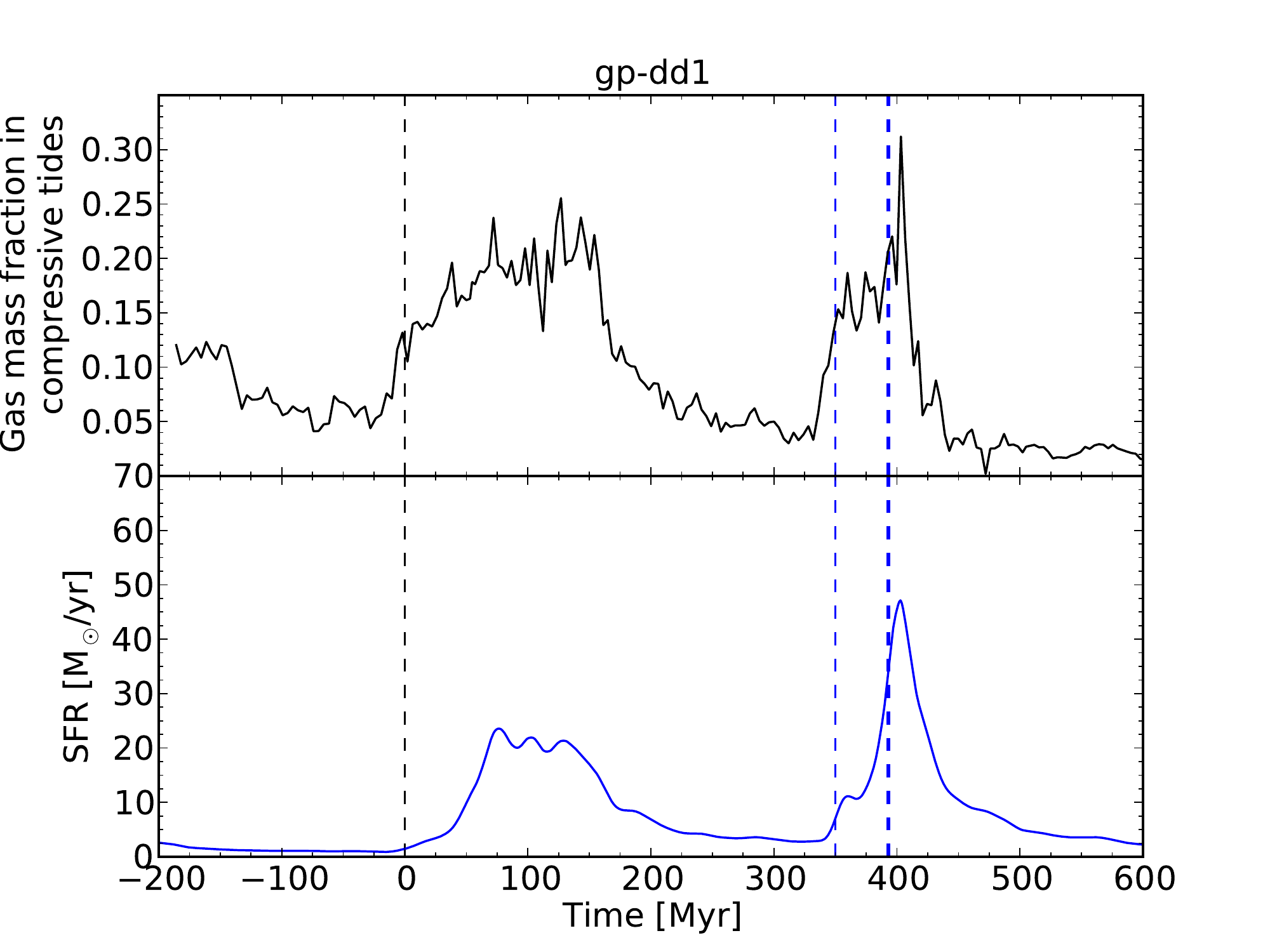}
\includegraphics[width=8.5cm]{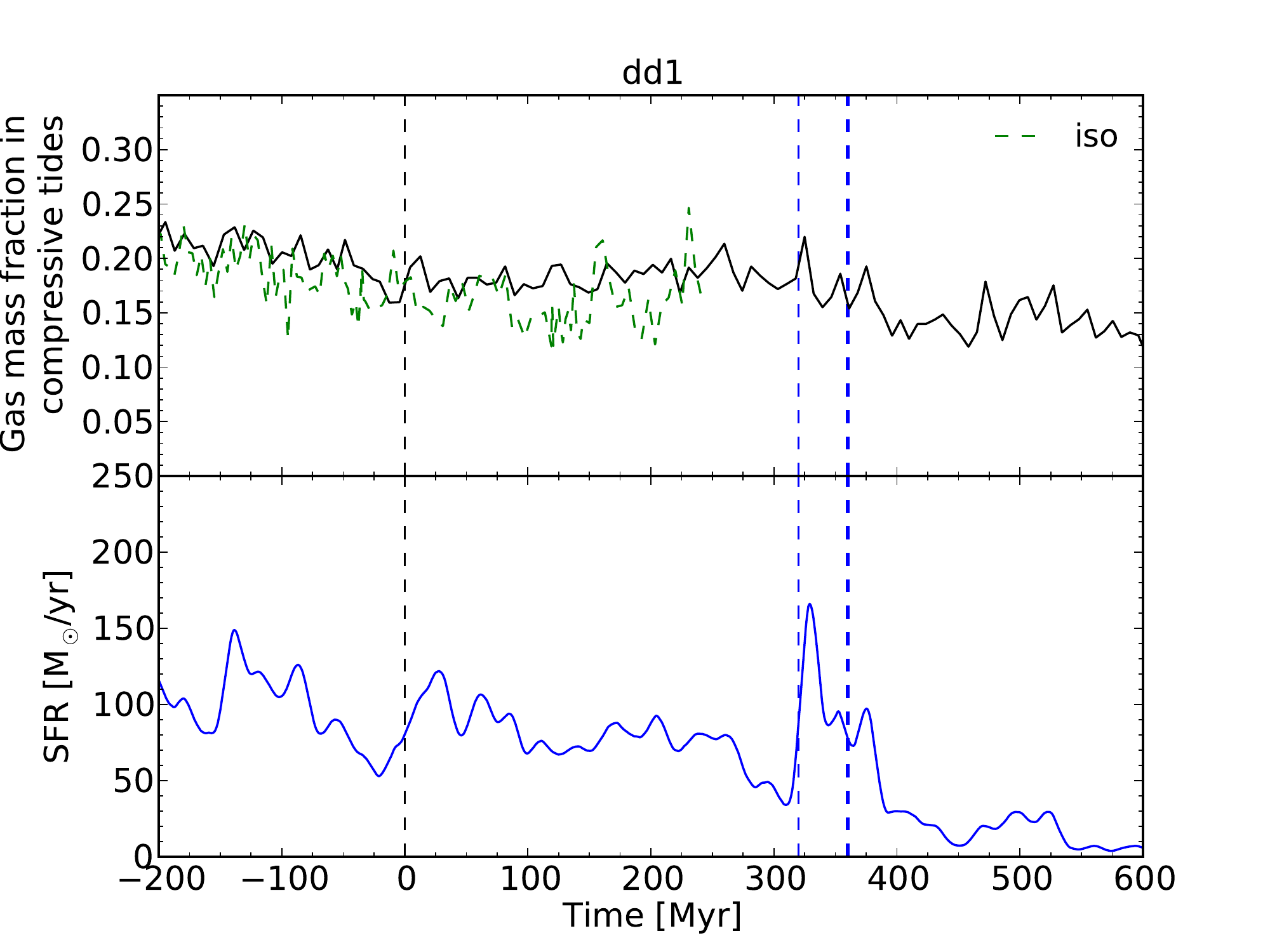}
\caption{Evolution of the gas mass fraction in a fully compressive tidal field (top panel) compared to the star formation rate (bottom panel) for the direct-direct encounter on orbit \#1 in the low gas fraction case (left) and high gas fraction case (right). The vertical dashed lines are the same as in Fig.~\ref{compSFR}.\label{tide}}
\end{figure*}

The increase in compressive turbulence is thought to be driven by the onset of fully compressive tides (see R14). On top of triggering compressive turbulence, fully compressive tides also reduce the Jeans mass and help gas fragmentation in star-forming clumps \citep{Jog13,Jog14}. Extended regions undergoing fully compressive tides are a common feature of galaxy encounters \citep{Renaud08, Renaud09}.

To measure the impact of compressive tides on our galaxies we compute the tidal tensor defined by its components $T_{ij} = - \partial_{i} \partial_{j} \phi$ using first order finite differences of the total gravitational force at the scale of 50~pc. The tidal field is compressive if the maximum eigenvalue of the tensor is negative. The method is the same as R14. Results for the low and high gas fraction runs on the Orbit \#1 are shown on Fig.~\ref{tide}.

We see that, in the low gas fraction case, the gas mass fraction in compressive tides increases from 7\% to 20\% during the pericentre passages. These values are similar than those obtained in previous simulations of galaxy interactions \citep[using {\it N}-body :][using AMR: R14]{Renaud08,Renaud09}.

In the high gas fraction case, the mass gas fraction in compressive tides is initially higher and tend to slowly and monotically decrease with time, with no significant changes induced by the pericentre passages.

\begin{figure*}
\includegraphics[width=4.2cm]{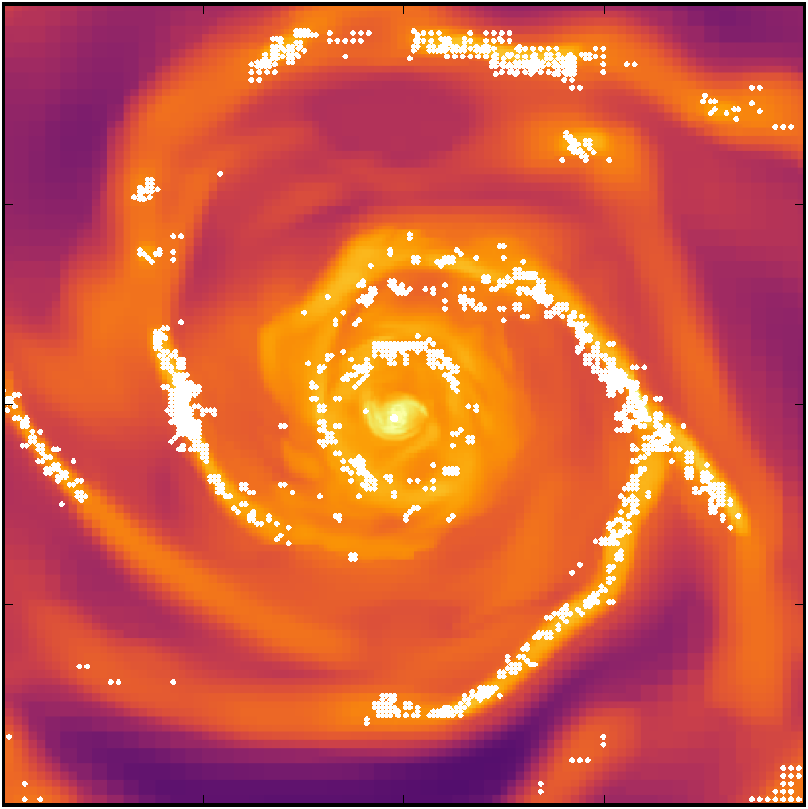}
\includegraphics[width=4.2cm]{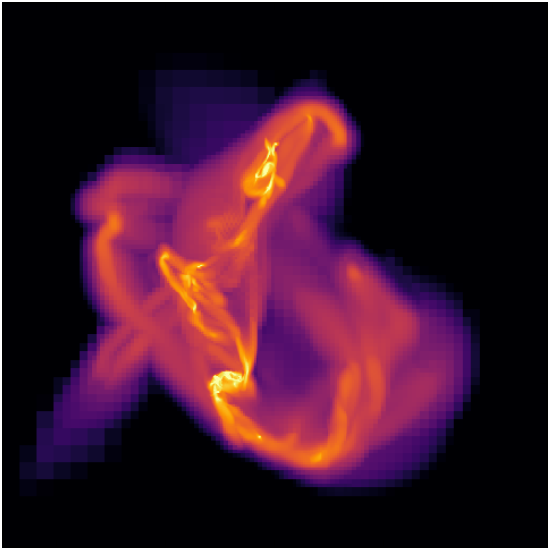}
\includegraphics[width=4.2cm]{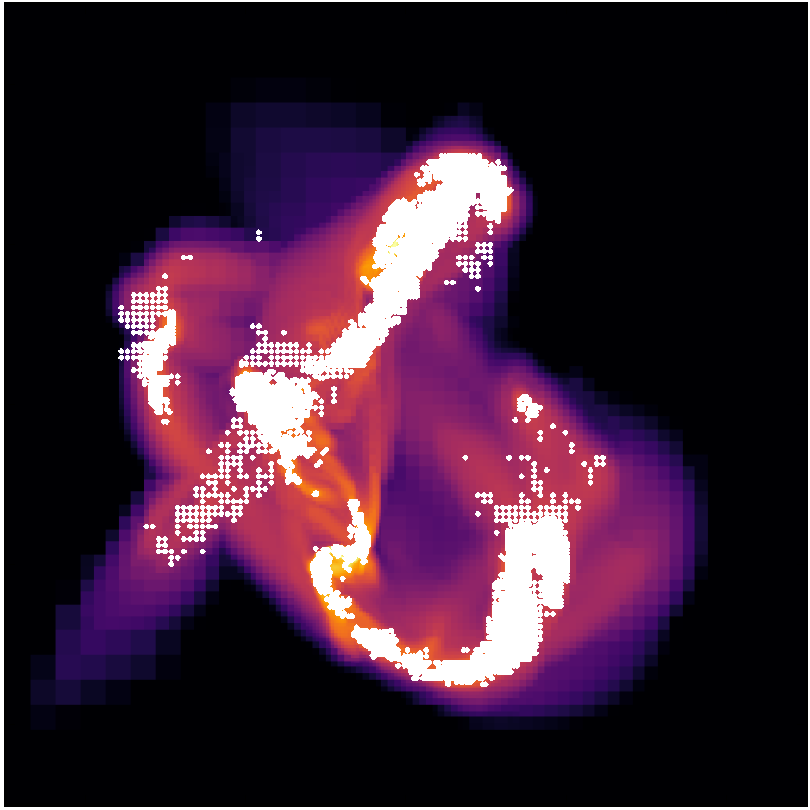}\\
\includegraphics[width=4.2cm]{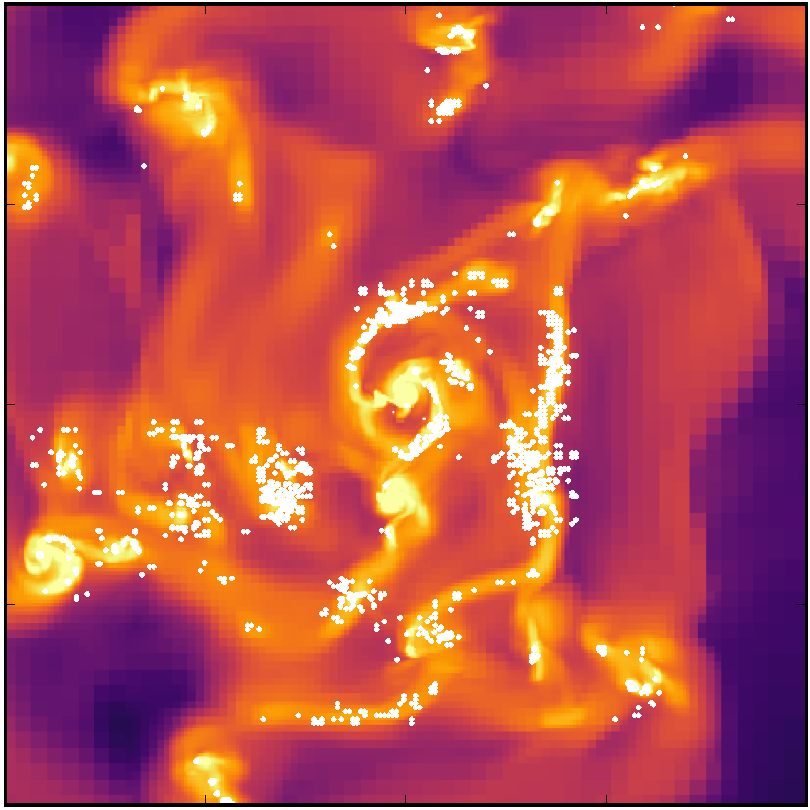}
\includegraphics[width=4.2cm]{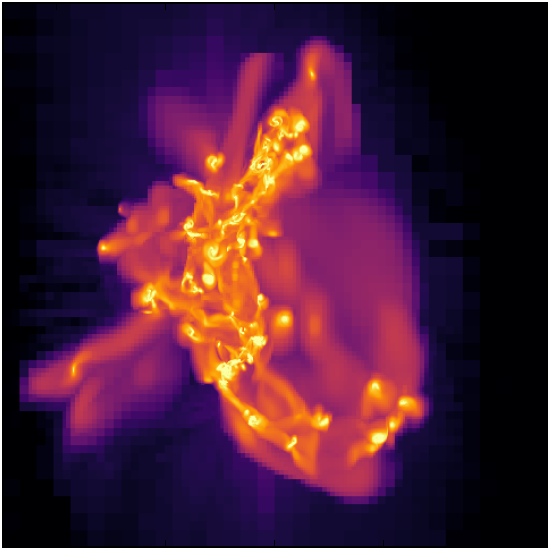}
\includegraphics[width=4.2cm]{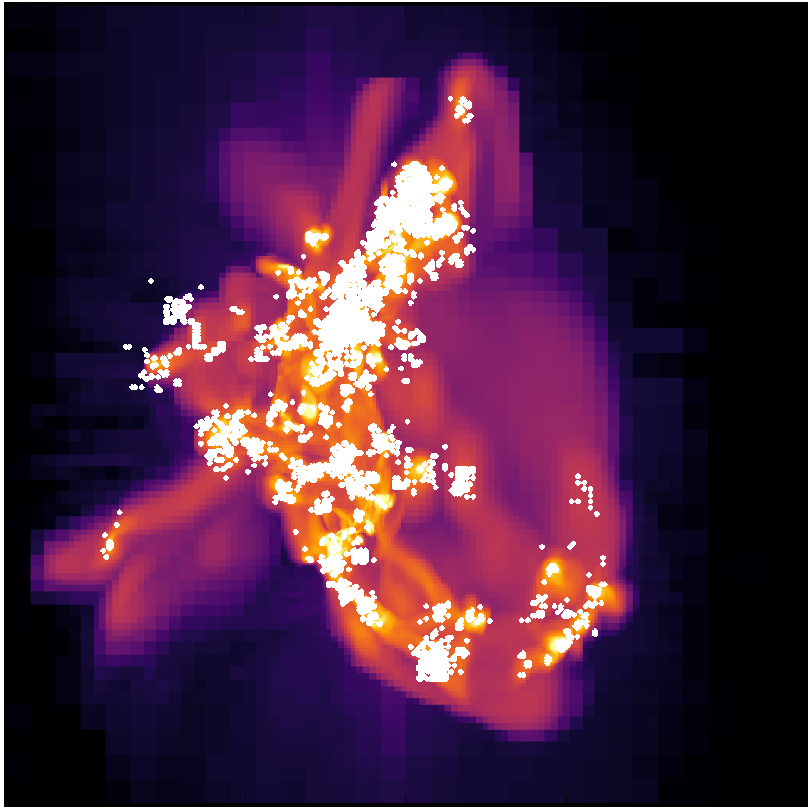}
\caption{Gas density maps for the gas poor (top panel) and gas rich (bottom panel) simulations. The left panels show the isolated galaxies. The central and right panels show the direct-direct interaction for orbit \#1 at t = 60~Myr. The maps span 10~kpc x 10~kpc for the isolated runs and 50~kpc x 50~kpc for the merger runs. The colormap is the same is in Fig.~\ref{isomap}. The white dots in the left and right columns show the location of the tidally compressive regions. \label{tide_map}}
\end{figure*}

This high but steady gas mass fraction in compressive tides results from the matter distribution of the galaxy. In Fig.~\ref{tide_map} we show the position of tidally compressive regions $\simeq$ 60~Myr after the first pericentre passage. In the low gas fraction case, we see that tidally compressive regions develop over extended regions: close to the galaxy nuclei, in the forming tidal tails and in the bridges between the two galaxies. In the high gas fraction case, the tidally compressive regions are mainly located inside the clumps. The gravitational potential is locally dominated by the highly concentrated clumps (see Section~\ref{morphology}), leaving the inter-clump medium in tidally extensive zones which makes the formation of any zone of compressive tides harder, and limits any increase in gas fragmentation during the interaction. 

This adds to the small enhancement of the turbulence, described in Section~\ref{turb}, which limits an increase in compressive turbulence, and thus a change in the density PDF and in the interaction-induced SFR.\\

In summary, our numerical simulations show that the clumpy nature driven by the high gas fraction has a strong influence on the central inflows, the gas turbulence and the compressive tides. These three physical processes are seen to be enhanced during low-redshift encounters and are thought to be responsible for the interaction-driven starburst in low redshift galaxies. In the high gas fraction cases, these three processes are already strong in isolated galaxies and are not further enhanced by interactions.

Our simulations hereby confirm the importance of these three processes in the increase of the SFR in gas-poor interactions and we claim that their weak enhancement for high gas fraction disks explains the relative diminished efficiency of gas-rich major mergers to trigger starbursts.

\subsection{No effect of saturation from feedback}

\begin{figure}
\includegraphics[width=9cm]{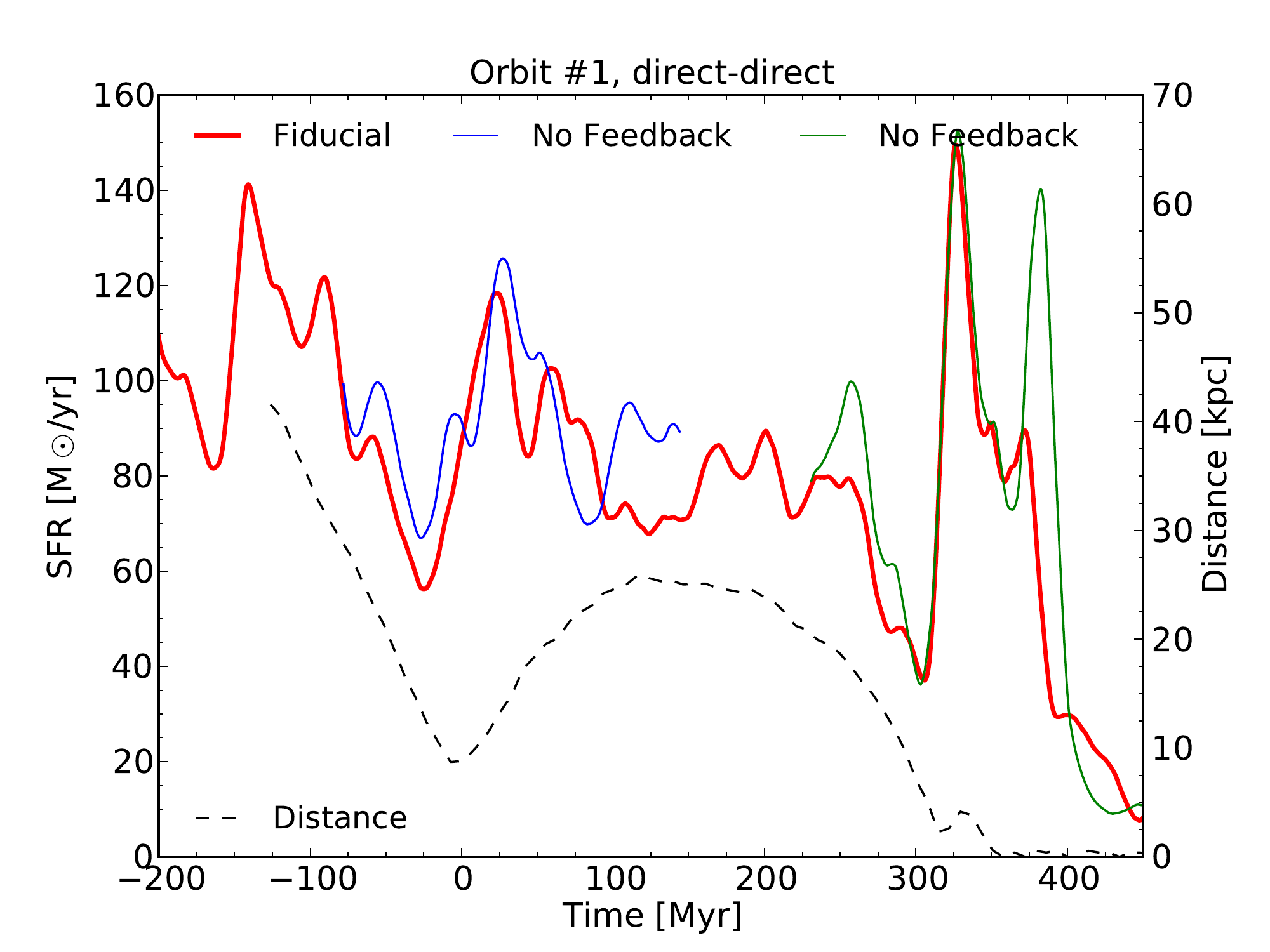}
\caption{Evolution of the SFR for the dd1 simulation and runs with turned off stellar feedback at different times. The red solid line shows the SFR of the fiducial run, dd1, presented in Fig.~\ref{compSFR}. The blue and green lines show the SFR resulting from the runs without feedback. They separate from the fiducial curve slightly before the time delay indicated in the text because of the curve smoothing. The black dashed line shows the distance between the two galaxies. \label{nofb}}
\end{figure}

The SFR of the high gas fraction disks in isolation is $\simeq 60$ times higher than for the low gas fraction disks, the net feedback energy from supernovae explosions and H{\sc ii} regions is therefore also much higher. One can wonder whether an increase of star formation is not self-regulated in high gas fraction disks because of the stellar feedback.

To test this hypothesis we restart the dd1 simulations, but with  all sources of feedback, which were presented in Section~\ref{code}, turned off. One should note that an absence of feedback for a long period of time would transform significantly the structure of the galaxies, as stellar feedback regulates the growth of gas clumps \citep{Bournaud10, Hopkins11b, Goldbaum16}. Shutting off feedback just before the investigated event, i.e the galactic encounters, ensures that we compare galaxies with the most similar structure as possible. Therefore, we shut off the feedback around 40~Myr only before the two moments when a rise of SFR is expected, namely the first pericentre passage and the coalescence. The time scale of dissipation of the turbulence being around 10~Myr \citep{McLow99} at the clumps scale, which is comparable to the free-fall time, we ensure that there is no more influence of the feedback on the gas turbulence at the expected starburst time.

The evolution of the SFR for the feedback case and no-feedback case is plotted in Fig.~\ref{nofb}. We see that there is a small increase in SFR, especially at coalescence, in the simulations without feedback, but the general behavior does not change: even without stellar feedback the interaction only induces a small starburst compared to the low gas fraction case. This weak influence of feedback on high gas fraction major mergers has already been seen in \citet{Bournaud11} and  \citet{Powell13}. 

Our simulations show that our feedback implementation, which allows a burst of star formation for low gas fraction major mergers, is not responsible for the weakness of the star formation enhancement in the high gas fraction case. The implementation of sub-grid models of stellar feedback in numerical simulations is a long standing issue. Energy outputs from SNe and stellar winds are theoretically rather well understood for most stellar populations \citep[see review by][]{Dale15}, but the details in the numerical implementations are important and can lead to significantly different results \citep[see e.g.][]{Hopkins12b, Agertz13}. A detailed study of the impact of the implementation of feedback on galaxy structure is beyond the scope of this paper.

On the one hand, our no-feedback simulations show that a weaker feedback than ours would not produce a strong enhancement of star formation either. On the other hand, a stronger feedback would increase the pre-interaction overall gas turbulence. As we have seen in Sect.~\ref{turb}, the interaction-induced increase of turbulence would be even smaller which would lead to an even weaker enhancement of the SFR. This shows that the weakness of merger-driven starbursts at high redshift (compared to low redshift cases) does not result from saturation by feedback in our models, and this result could only be stronger if real feedback had more important effects in high-redshift galaxies.


\section{Discussion}

\subsection{Comparison with previous simulations}

Our simulations show that the high gas fraction major mergers are less efficient than low gas fraction major mergers to trigger starbursts. Here we compare our results to previous studies of clumpy gas-rich major mergers:

\begin{itemize}

\item The first simulations of gas rich ($f_{\mathrm{gas}} > 50\%$) galaxy mergers could not resolve the impact of turbulence on star formation, \citep{Springel05, Hopkins06, Cox06, Lotz08, Moster11}: their disks are less unstable than ours and tend to form spirals instead of clumps. They resolve the star formation enhancement due to the increase of nuclear inflows through interaction-driven gravitational torques, but miss the increase of gas fragmentation and the initial, clump-driven, nuclear inflows, which make them to overestimate the relative increase of SFR.

\item \citet{Bournaud11} have run the first numerical simulations of gas rich clumpy galaxies. On their three orbits, they saw an increase of the SFR of up to a factor 10 (from 200~M$_{\odot}$~/yr to peaks of 2000~M$_{\odot}$~/yr) for more than 200~Myr. It results in a higher increase of SFR than in our sample, but still below usual low gas fraction major mergers. One should note that they do not use a cooling function, but a barotropic equation of state \citep[presented in][]{Bournaud10}, which tends to over-estimate the gas fragmentation. 

\item \citet{Hopkins13} present a  suite of parsec-resolution galaxy collision simulations, with one high-redshift type galaxy major merger ($f_{\mathrm{gas}} = 50\%$) and a feedback implementation which systematically destroys gaseous clumps. Their Fig.~9 shows that the enhancement of star formation is indeed significantly smaller and has a $\simeq$ 10 times shorter relative amplitude than their Milky Way type galaxy major merger. Their starburst duration is similar to that hat obtained in our simulations: $\sim$ 100~Myr and $\sim$ 1~Gyr for respectively the high and low gas fraction cases. They did not study this weak enhancement in details, as they changed several parameters between the low and high redshift runs (halo mass, the disk compactness and velocity profile).

\item In the merger sample of \citet{Perret14} no simulation shows any increase of the star formation during the interaction. They stressed that one explanation for this could be due to their implementation of feedback: they set a density-independent cooling time for the gas heated by the SNe of 2~Myr, which prevents very dense region in star-forming regions to form stars again during this time. One should note that they did not test their feedback on low gas fraction major mergers, contrary to the present study, so that it is uncertain whether this implementation would allow them to trigger starbursts in low redshift conditions.
\end{itemize}

The fact that high gas fraction major mergers trigger a smaller enhancement of star formation than their low gas fraction equivalent already was already seen in previous studies, using various numerical methods and feedback implementations. With this suite of simulations we propose that the variation in gas fraction only can explain the mildness of this enhancement.

\subsection{The cosmological context}
\label{infall}

\subsubsection{Gas refueling}

\begin{figure}
\includegraphics[width=9.5cm]{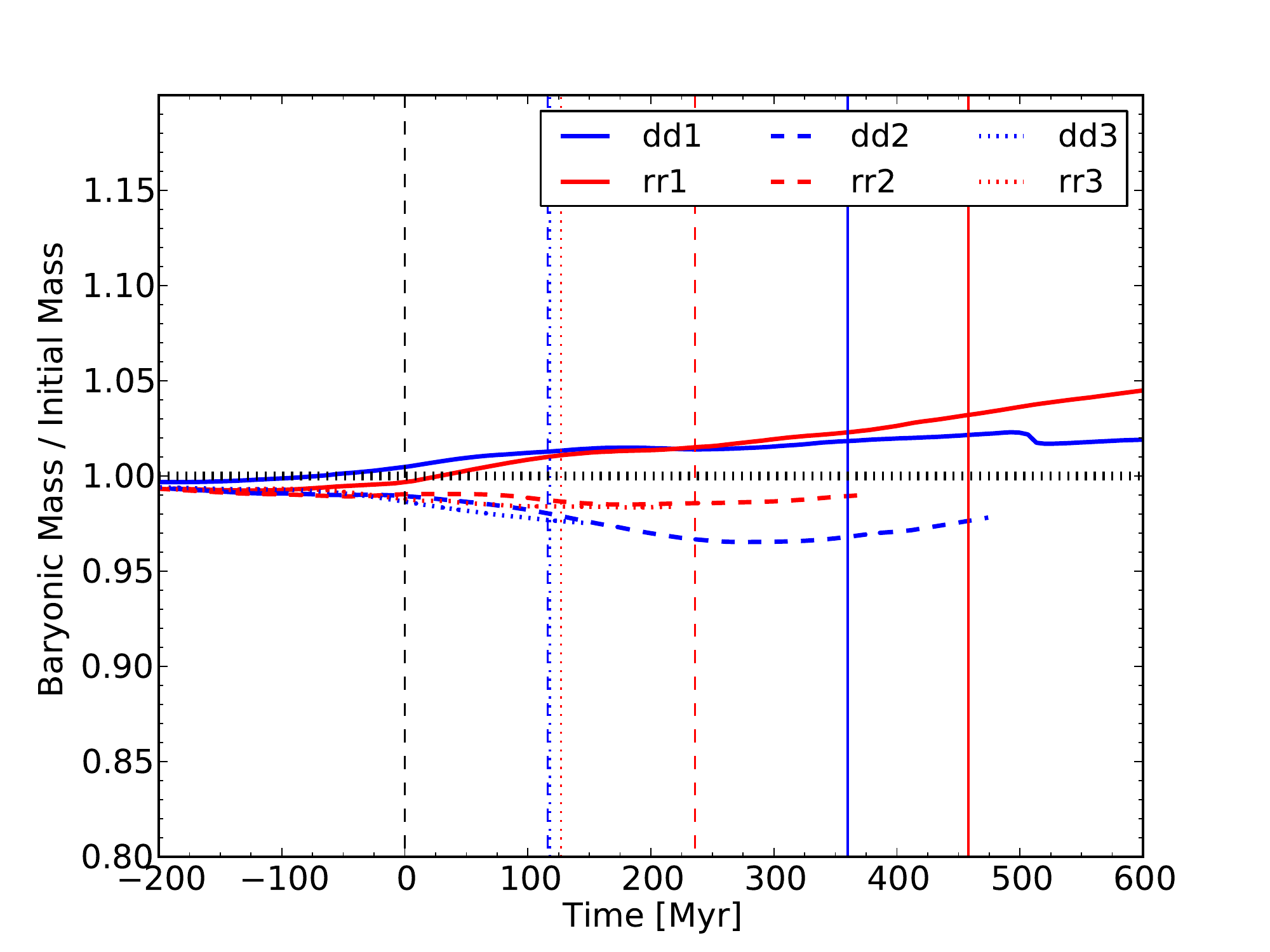}
\includegraphics[width=9.5cm]{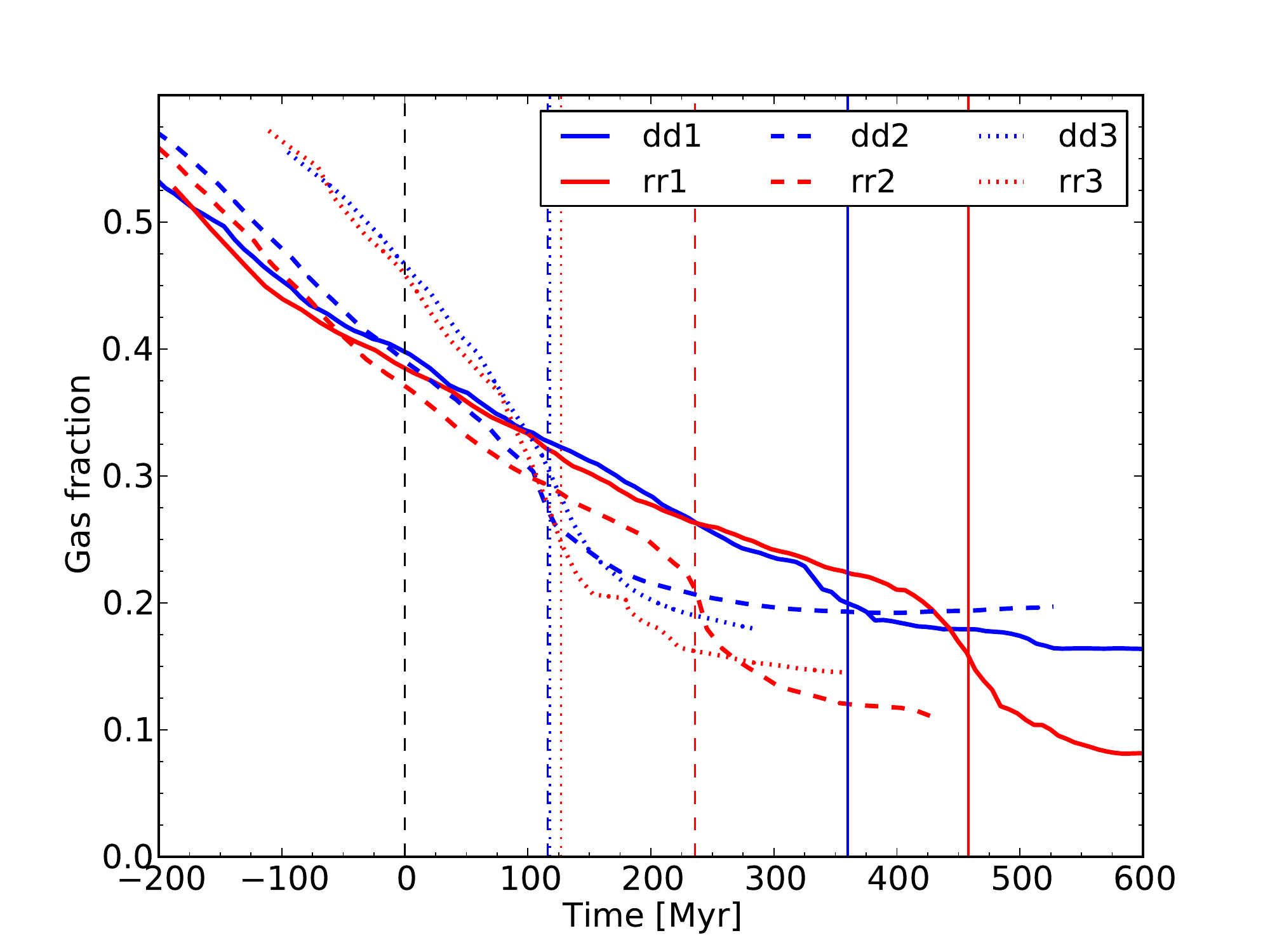}

\caption{ {\it Top:} Amount of baryonic matter ( $  M_{\mathrm{gas}} + M_{\mathrm{stars}} $) renormalized to the initial baryonic mass: 2 $\times$ 57.2 $\times$ 10$^{9}$ M$_{\odot}$ (see Table~\ref{galaxy}). The vertical dotted line shows a ratio of 1.  {\it Bottom:} Gas fraction evolution during the simulation for six gas-rich orbits. For both panels, only gas with density above 10$^{-2}$~cm$^{-3}$ is considered, in order to only count the gas present in the galaxies. All curves are shifted in time, such that $t=0$ is the time of first pericentre for each orbit. The small  coloured vertical lines show the time of coalescence for each orbit, as labeled in the legend.  \label{gasfrac}}
\end{figure}

Our galaxy pairs are not refueled by cosmological gas infall. The presence of cosmological cold gas infall on galaxies, inferred from cosmological simulations \citep{Dekel09a, Dekel09b} but not yet observed, is thought to be the main channel of galaxy mass growth at high redshift, when major mergers would contribute to only one third of the mass growth \citep{Keres05, Ocvirk08, Brooks09}.

Even though no cosmological infall is considered in our work, the galaxies are initially located in a box with an inter-galactic medium (IGM) of background density 2~$\times$10$^{-7}$ cm$^{-3}$. This gas can cool down and contribute to the baryonic mass budget of the galactic disks. 

In the top panel of Fig.~\ref{gasfrac} is shown the quantity of baryonic matter in the galaxies (stars and gas above 10$^{-2}$ cm$^{-3}$), normalized to the initial baryonic mass of 2 times  57.2 $\times$ 10$^{9}$ M$_{\odot}$ (see Table~\ref{galaxy}). We see that the baryonic mass of the galaxy pair does not change by more than $4\%$ throughout the simulations. These variations are either due to stellar outflows, which can bring some gas to densities lower than our threshold of 10$^{-2}$~cm$^{-3}$,  or to the cooling of the IGM gas onto the disk. We see that these two processes do not affect our mass budget.

A first drawback of not adding gas accretion in our simulations is that the cold gas mass reservoir available to form stars is not replenished by these cold flows. For instance, in our two simulations on orbit~\#1 -- the less bound one for which coalescence happens after more than 300~Myr from the first pericentre -- the long term evolution of the SFR in Fig.~\ref{SFR} is seen to be slowly decreasing, as a the consequence of gas consumption.

In the bottom panel of Fig.~\ref{gasfrac} is shown the gas fraction evolution during the gas-rich simulations on orbits \#1, 2 and 3. We see that the gas is steadily consumed before the first pericentre, at $t = 0$, where the gas fraction is still at around $40 \%$ or more. Before the coalescence the gas fraction is above $25 \%$ for orbits \#1 and 2, and above $40 \%$ for the faster orbit \#3. Just before coalescence, the faster gas consumption through star formation leads to a rapid drop in the gas fraction. 

We see that for the orbit \#3 the effect of gas consumption is relatively weak as the gas fraction is still above $40 \%$ at the time of the coalescence. The fact that the SFR for these orbits do not show much difference in behavior from orbit \#1 comforts us in concluding that this would not have an important effect on the SFR of our simulations.

However, on longer timescale, gas accretion would impact the star formation in the merger remnant. Indeed, in Fig.~\ref{ks} we see that after the coalescence, the SFR of the gas-rich merger drops well below what is expected from the remaining gas fraction. This is due to the formation of a massive stellar spheroid, dispersion dominated, which stabilizes the gas and prevent it to form stars. This process is called morphological quenching \citep{Martig09}. Recent simulations of high gas fraction galaxies surrounded by hot gaseous haloes, presented by \citet{Athanassoula16}, show that a disk of gas can form through gas accretion inside the spheroid core. This disk will be bar-unstable and form stars. 

At $z=2$, the mass inflow from cosmological accretion is of the order of 10~M$_{\odot}$/yr \citep{Dekel09a}. \citet{Martig09} have shown that, with this accretion rate order of magnitude and without any other merging event, several Gyr are necessary to reform a star-forming disk in a spheroidal major merger remnant, which is the same order of magnitude of formation of a disk found by \citet{Athanassoula16}.

\subsubsection{Turbulence from cosmological infall}

A second effect is that we neglect the increase of turbulence from the cosmological accretion. The infalling gas does not settle immediately into a thin disk but tends to stir the existing gas disk \citep{Elmegreen10, Gabor13}, which has for net effect to increase the velocity dispersion and the disk height. The SFR is thought to be regulated by this constant fueling of turbulence onto the disk, until $z=2$ when this process stops dominating the turbulence budget \citep{Gabor13}.

In Section~\ref{turb} we claim that the merger does not increase the turbulence, because it is already high. Turbulence pumped through cosmological infall would make it even harder for the interaction to further increase the turbulence. Therefore we argue that this process would not change our conclusions.

\subsubsection{Galaxy compactness and orbits}
\label{diff}

A last effect is that high redshift galaxies are more compact than low redshift galaxies, for the same mass, typically  a factor $\simeq$ 2 \citep{vanderWel14, Ribiero16}. In this study we neglect this effect to focus on the difference in gas fraction only. However, our galaxy model has an intermediate compactness, between low and high redshift conditions.

Furthermore, for equal halo mass, cosmological simulations expect the pericentre distance to be 25\% lower at $z=2$ than at $z=0$ \citep{Wetzel11}. All dynamical timescales are then shorter and the merging process is faster. This could result in more starburst-favorable orbits.

The orbit \#2 (resp. \#3) has impact parameter and relative velocity respectively 15\% (resp. 30\%) lower than orbit \#1, which makes it closer to typical $z=2$ orbits. We see in Fig.~\ref{SFR} that these three orbits share a comparable increase in SFR. This hints for a low dependency of the orbit parameter on the weakness of the interaction-driven star formation in high gas fraction disks.

\subsection{Application to local gas-rich dwarf galaxies interactions}

Local dwarf irregulars (dIrrs) have f$_{\mathrm{gas}}$ up to 90\% , contain clumps and are relatively turbulent with a ratio of their circular velocity to velocity dispersion, $V_{\mathrm{circ}}$ / $\sigma$, around 6 \citep[see e.g.][and references therein] {Lelli14}, which is of the same order than massive high redshift galaxies. Our reasoning, which is based on the high gas fraction and highly turbulent high redshift galaxies, might thus also apply to dIrrs. However, \citet{Stierwalt16} recently claimed that interactions between dIrrs can significantly increase their SFR, that is more than 5 times above the main sequence.

It should be noted that, if the gas fraction of dIrrs is very high, it is mainly due to a diffuse H\textsc{i} envelope, which does not participate actively to the star formation. This envelope can however cool and be accreted onto the central region \citep[see e.g.][]{Elmegreen16b}. 

A second important difference with respect to high redshift massive galaxies is that isolated dIrrs do not show signs of instability-driven inflows, as they do not have important nuclear concentration. For the gas to be funneled towards the nucleus, the work of the gravitational torques from the gaseous clumps, which scale with their mass ($M_{\mathrm{clump}}$), must compensate the galactic rotational kinetic energy of the gas $\propto$ $v_{\mathrm{circ}}^{2}$, which results in an inflow timescale $\propto$  $v_{\mathrm{circ}}^{2}$ / $M_{\mathrm{clump}}$.

Both dIrrs and high redshift massive galaxies show clumps of typical masses of a few percent of the galaxy mass, that is $\simeq$ 10$^{7}$~M$_{\odot}$ \citep{Elmegreen12}, and rotation velocities of about 50~km/s \citep{Lelli14}, against 10$^{9}$~M$_{\odot}$ clumps \citep{Elmegreen09c} and 200~km/s rotation velocities for massive high-redshift galaxies. The inflow timescale is then a factor 6 longer for dIrrs. The fact that the relative torquing of the gas is much less efficient in dIrrs makes the central inflows driven by the interaction between dIrrs relatively more important and can thus more easily ignite a starburst. 

The more diffuse location of the gas in dIrrs, and the low efficiency of instability-driven central inflows therefore limit the extension of our results from massive high redshift galaxies to dIrrs.

\subsection{Starbursting galaxies observed at high redshift}

In Fig.~\ref{compSFR} and \ref{SFR} we see that interactions do not strongly enhance the SFR of high gas fraction galaxies: from around 120~M$_{\odot}$/yr (60~M$_{\odot}$/yr for each galaxy), our most actively star forming galaxy barely reaches 350~M$_{\odot}$/yr during 50~Myr. Observations of sub-millimeter galaxies (SMGs), report SFRs up to several 1000~M$_{\odot}$/yr \citep{Barger98, Tacconi06}. Observed SMGs are found both on the main sequence and the starburst sequence of the $M_{\star}$-SFR \citep{deCunha15}. There are also claims that SMGs form the high-mass end of the main-sequence \citep{Koprowski16, Michalowski16}. 

Recent numerical simulations presented in \citet{Narayanan15} show that the regime of SMGs galaxies can be achieved without major mergers, but rather by continuous gas infall through accretion of small haloes onto a very massive group halo ($\simeq 10^{13}$~M$_{\odot}$) which may be populated by several galaxies. On top of that, \citet{Wang11} show that the relatively high abundance of SMGs might result from the blending of sources that are not necessarily spatially associated, which agrees with a not high enough number of major merger to account for the observed SMG abundance \citep{Dave10}. 

These studies do not exclude major mergers as being a cause of a part of the observed SMGs. However, one should note that the progenitors of such galaxies are significantly more massive than our simulated disks ($M_{\star}$ $\simeq 10^{11} $M$_{\odot}$), i.e five times higher than our galaxy models. Therefore, we  cannot conclude on the possibility of part of the SMGs being triggered by high-redshift major mergers.


\section{Conclusions}

Observations of star-forming galaxies over a wide range of redshifts ($z=4$ to $z=0$) have shown a main sequence on the $M_{\star}$--SFR relation, with a number of outliers galaxies showing a higher specific SFR, which correspond to the starbursting galaxies. The proportion of these outliers stays relatively constant ($\simeq 2\%$) with time, with no clear dependence with redshift. In the local Universe, all these outliers are merging galaxies. Since the merger rate is thought to increase with redshift, high redshift major mergers seem to be less efficient to produce starbursting galaxies than low-redshift ones.
We present a suite of parsec-resolution hydrodynamical simulations to test this hypothesis. We have run simulations of typical low-redshift and high-redshift galaxy pairs, changing only the gas fraction, from 10~\% to 60~\%, between the two models. 

We see that for the same orbit, when the SFR of the low gas fraction goes up by a factor of more than 10 during several hundred million years, the SFR of the high gas fraction case only mildly increases, and only at the coalescence. In particular we show that the high redshift type galaxies would not, or for a short amount of time ($\simeq$ 50~Myr), be considered as starburst galaxies on a Schmidt-Kennicutt diagram. 

Even in the absence of stellar feedback no higher burst of star formation is seen. We reject feedback saturation as a cause to the inefficiency of the interaction to trigger a strong enhancement of star formation.

At coalescence, most of the star formation happens in the central region of both galaxies, typically in the central kpc. This starburst is fueled by central mass inflows caused by the interaction, which add up to the mass inflow processes already at play in the isolated disks. This makes the coalescence phase less efficient in triggering a starbursting galaxy in the high gas fraction case than the low gas fraction case.

\citet{Renaud14} have shown a correlation between the increase in compressive turbulence and the onset of the interaction-driven starbursting mode of star-formation. Our high-redshift type galaxy interactions show only a mild increase of turbulence. High gas fraction clumpy disks are already quite turbulent ($\sigma \simeq$ 40 km/s) and it is harder to increase the velocity dispersion through release of gravitational energy when the initial velocity dispersion is already high. 

On top of this smaller increase of the gas turbulence, the clumpy morphology also imposes an extensive tidal field on the inter-clump regions. These two processes limit the increase in compressive turbulence, and thus the increase in star formation rate during the interaction.

This demonstrates that the high gas fraction of the galaxy progenitors can strongly reduce the star formation enhancement in galaxy interactions and major mergers.

\section*{Acknowledgements}
The authors thank Vianney Lebouteiller and Damien Chapon for very useful discussion and the anonymous referee for constructive comments which helped to improve this paper. Simulations were performed at TGCC (France) and as part of a PRACE project (grant ra2540, PI: FR) and GENCI project (grant gen2192, PI: FB). FR acknowledges support from the European Research Council through grant ERC-StG-335936.




\bibliographystyle{mnras}
\bibliography{draft} 




\appendix

\section{Study of spatial resolution}
\label{resolution}

\begin{figure}
\includegraphics[width=8cm]{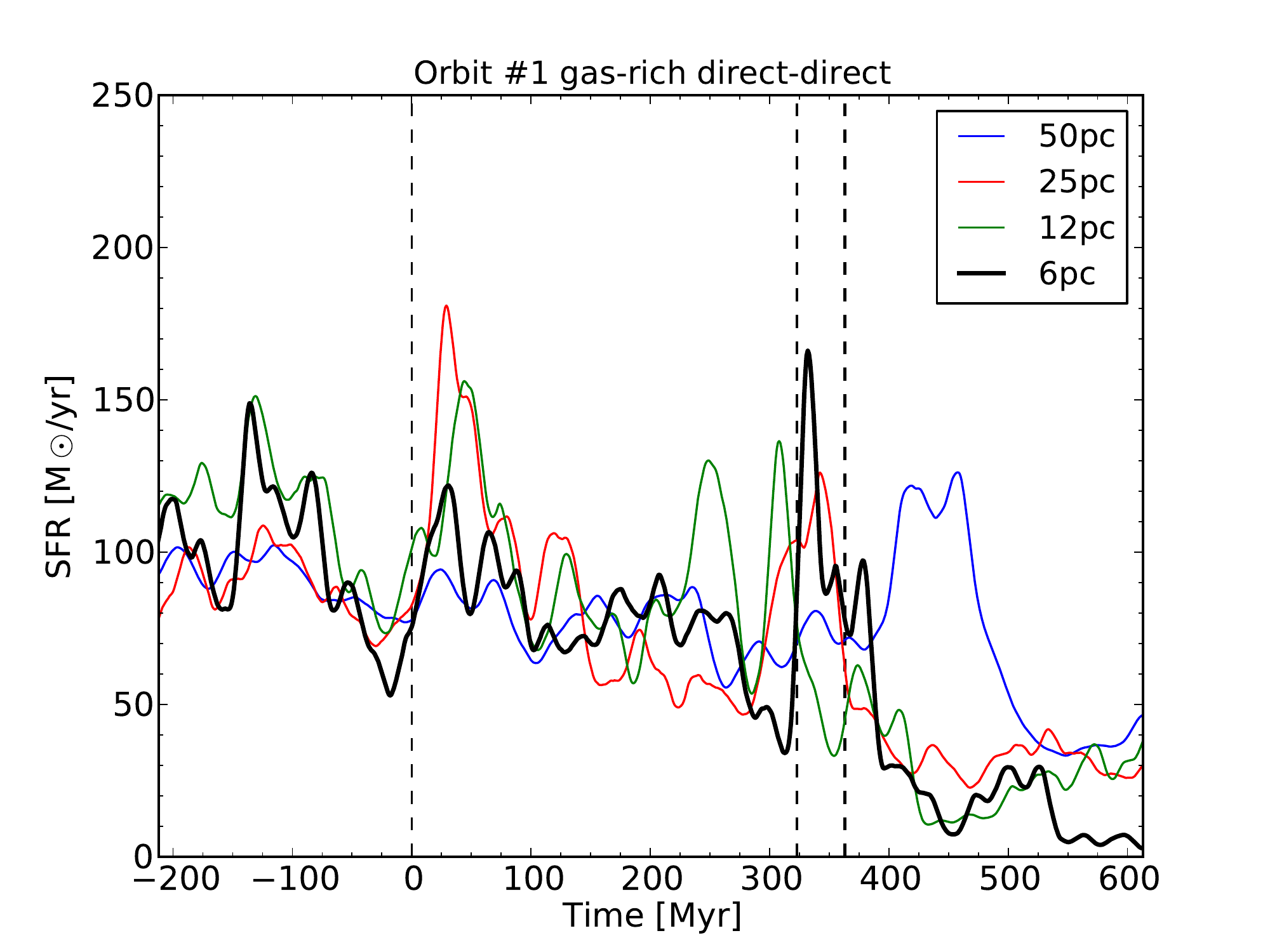}
\includegraphics[width=8cm]{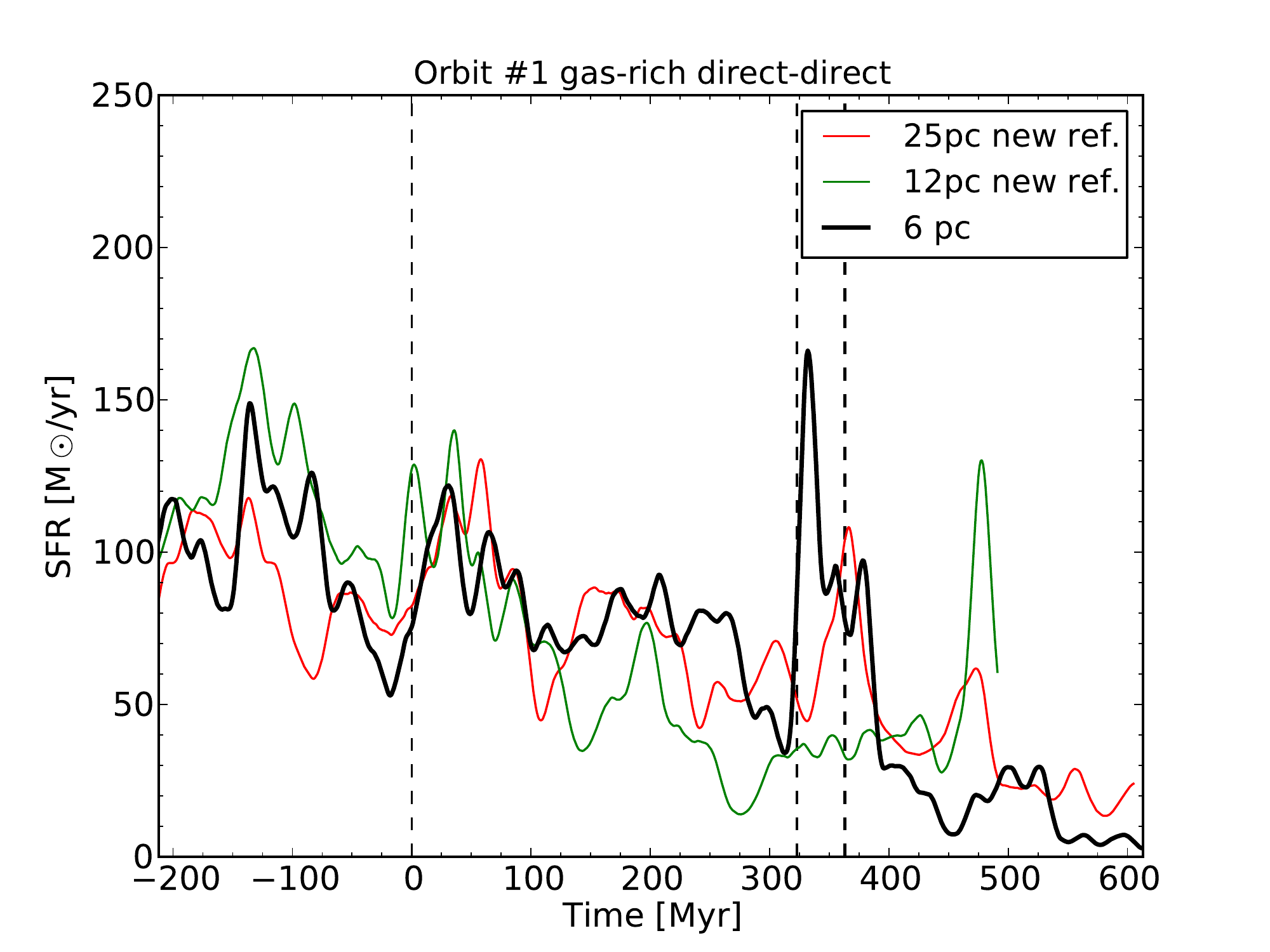}
\caption{{\it Top}: SFR during the interaction for the dd1 simulation at different maximum spatial resolution. {\it Bottom}: Same as in the top panel, with a different refinement strategy for the new refinement labeled simulations. The black dotted lines show the time of pericentre passages and coalescence of the dd1 simulation at 6~pc resolution. \label{reso}}
\end{figure}

In order to test the impact of the resolution on star formation of our high-redshift simulations, we run the dd1 simulation with different resolution both in space and in mass. 

In the upper panel of Fig.~\ref{reso} is shown the SFR for the dd1 simulations for different highest spatial resolution, from 50 to 6~pc. In the lower panel of Fig.~\ref{reso} is shown the SFR for the dd1 simulations for different highest spatial resolution, using a different mass resolution. For the two new refinement simulation, the mass needed to activate the next refinement level was divided by 8 for cells smaller than 100~pc. Therefore, all cells that were refined to 50~pc in the non-new refinement simulations, are directly refined to 25~pc in the new refinement case, and so on for further refined cells.

The stochasticity of the SFR and the fact that each simulation is run with a different $\rho_{0}$ calibration induces some variations in the pre-merger SFR. The time of the final coalescence can change with the resolution, as the resolution will impact several processes governing the interaction. For instance, an other resolution can modify how the angular momentum is driven out of the inner system through the tidal tails and/or impact the dynamical friction.

In Fig.~\ref{reso}, there appears to be no significative trend with the resolution. In particular, no simulations show any onset of a massive star formation burst during the interaction. We therefore see that the resolution has little influence on the evolution of the SFR during the interaction, which legitimates our use of a resolution of only 12~pc for the simulations run with Orbit \#2 and Orbit \#3.


\bsp	
\label{lastpage}
\end{document}